\documentclass[aps,prb,twocolumn,nopacs,amssymb]{revtex4}

\usepackage{amsmath}
\usepackage[dvips]{graphicx}
\usepackage{dcolumn}
\usepackage{bm}
\def\be{\begin{equation}}
\def\en{\end{equation}}

\def\p{\partial} 
\newcommand{\av}[1]{\langle{#1}\rangle}

\def\gs{\gtrsim}

\newcommand{\bi}[1]{\mbox{\boldmath$#1$}}
\def\p{\partial}
\def\bea{\begin{eqnarray}}
\def\ena{\end{eqnarray}}
\def\a{_{\alpha\beta}}

\def\aP{\stackrel{\leftrightarrow}{\Pi}}
\def\ab{{\alpha\beta}}
\def\ih{{\rm ih}}

\begin{document}


\title{Theory of Shear  Modulus  in Glasses}

\author{ Akira Onuki$^1$ and Takeshi  Kawasaki$^2$ }
\affiliation{
$^1$Department of Physics, Kyoto University, Kyoto 606-8502, Japan\\ 
$^2$Department of Physics, Nagoya University, Nagoya 464-8602, Japan\\
}

\date{\today}

\begin{abstract}
We construct a linear response theory of 
 applying shear deformations from  boundary walls in the film 
geometry in Kubo's theoretical scheme.  Our method is applicable to any solids 
and  fluids.  For   glasses, 
we assume quasi-equilibrium around a fixed inherent 
state. Then, we   obtain  
linear-response  expressions   for any variables including 
the stress  and the particle displacements, even though 
the glass interior is  elastically inhomogeneous. 
In particular,  the shear modulus   
can be expressed in terms of the correlations  
between the interior stress and the forces from the walls. 
It can also be expressed in terms of the inter-particle 
correlations, as has been shown  in the previous literature. 
Our stress relaxation function 
includes the effect of the boundary walls 
and can be used for inhomogeneous flow response.  
We show the presence of 
long-ranged, long-lived  correlations among  the fluctuations of the forces 
from the walls and the  displacements of all the particles in the cell. 
We confirm these theoretical results numerically 
in  a two-dimensional model glass.
As an application, we 
  describe   propagation of transverse sounds  after  boundary wall motions  
using these time-correlation functions 
We also find resonant sound amplification  
when the frequency of an oscillatory shear 
approaches that of the first transverse sound mode. 
\end{abstract} 


  

\maketitle


\section{Introduction}

In  glasses,    the 
structural relaxation   becomes 
 exceedingly  slow  at low temperature $T$. 
The shear modulus $\mu$   is 
 then well-defined  for  small    deformations,  
 though plastic events easily take place  with increasing  
the applied strain\cite{Liu,Lacks,Maeda,Yama1,Kawasaki,Malo2,Barr3}. 
It is of great interest how $\mu$ in glasses   depends on  the 
disordered particle configuration. 
On the other hand, in  crystals, the microscopic expressions 
for the elastic moduli can    be derived  under a homogeneously  applied 
strain (or stress) in equilibrium\cite{Hoover,Ray,Lutsko,Pablo,Hess}. 
Such expressions  are  composed of a positive 
affine   part and a negative nonaffine part, where the latter arises from 
 the correlation of the stress fluctuations. 
If the moduli are  homogeneous, they    can  be 
related to  the variances of the thermal strain (or stress) 
fluctuations divided by $k_BT$\cite{Rahman,Binder,Gusev}.

In glasses, $\mu$ is     expressed  in 
 the same  form as  those in  
crystals  in terms of the particle 
positions on timescales without plastic 
events\cite{Malo,Malo1,Evans,Wil,Yoshi,Ilg,Barr1,Barr2,Saw,Wit,Wit1,Teren,Za}.   To derive this expression, 
  Maloney and Lema\^itre\cite{Malo,Malo1} 
examined   the local minima of the potential energy 
under  constraint of a fixed mean strain 
in the periodic boundary condition.
Remarkably, the  local values of $\mu$    
 exhibit  mesoscopic inhomogeneity\cite{PabloP,Miz1}. 
In fact, in  glasses, 
 the displacements (and suitably    defined strains) in  glasses    
 are highly heterogeneous  on mesoscopic scales  
under  shear\cite{Barr1,Barr2,Malo2,Yama1,Kawasaki,Liu,Maeda,Barr3}. 

The linear response theory  in statistical 
mechanics has a long history\cite{Hansen,Onukibook,Zwan}. 
On the basis of Onsager's theory\cite{Onsager},    
 Green\cite{Green} derived time-evolution equations 
for {\it gross variables}, 
which was rigorously  justified by Zwanzig\cite{Zwan}. 
Green  then  expressed  the transport coefficients 
in fluids such as  the viscosities and the thermal conductivity 
 in terms of the time-correlation functions of the stress 
and the heat flux, respectively.   These expressions    also followed   
from the relaxation  behaviors   of the 
time-correlation functions  of   hydrodynamic 
variables\cite{Hansen,Kada,Zwan}.
   Kubo\cite{Kubo}  studied  
linear response to  {\it mechanical  forces}, 
for which   the  Hamiltonian consists  of the unperturbed  one  $\cal H$ 
and  a small time-dependent perturbation  as  
\be 
{\cal H}'= {\cal H}- \gamma_{\rm ex} (t) {\cal A}. 
\en   
Here,   $\gamma_{\rm ex}(t)$ is  an   applied force 
 and ${\cal A}$ is its  conjugate variable.    
Thus, there  was  a conceptual  difference 
between the  approaches to 
thermal and  mechanical disturbances.
 
In this paper, we set up  a Hamiltonian in the form of Eq.(1) 
 for slight motions of the  boundary walls  in  the film geometry, 
where   the film thickness $H$ is  much longer  than the particle sizes.
Here,  $\gamma_{\rm ex}$  is the  {\it mean}  shear strain,     
and $\cal A$ is  given by  
$H({F}^x_{\rm bot}-{F}^x_{\rm top})/2$, 
where  ${F}^x_{\rm bot}$ and ${F}^x_{\rm top}$ 
are the  tangential forces from the bottom and top 
walls to the particle system. For glasses,     we 
can examine   linear response    for any variables assuming 
quasi-equilibrium   around  a fixed inherent state
\cite{Malo,Malo1,Evans,Wil,Yoshi,Ilg}. This is 
justified   while jump motions do not occur among 
different inherent states\cite{Sastry,Heuer,Lacks,Harro1}. 
For  liquids, our theory yields Green's 
 expression for the shear viscosity with  
Kubo's method in the low-frequency limit. 
It  can further  be used to analyze  linear response   
in fluids  near a moving    wall\cite{slip}.

In our theory,  $\cal A$ in Eq.(1) 
can be  expressed in terms of the particle positions 
near the walls. However, for nonvanishing   $\mu$,   
 the    fluctuations of $\cal A$ 
 are   significantly correlated  with  those of 
all the particle displacements ${\bi u}_i$  in the film  due to 
the large factor $H$ in its  definition.  
We shall even find a correlation  between    the   fluctuations of 
 ${ F}^x_{\rm top}$ and ${ F}^x_{\rm bot}$   proportional to $\mu/H$.  
On the other hand, in infinite glasses $(H\to \infty)$,  
 the stress pair correlation   decays   
algebraically in space (under the periodic boundary condition in 
simulations)\cite{Lema,Fuchs,Seme,Harro1,Egami}.
We mention a  similar effect in  polar fluids, where the 
 polarization pair correlation  is 
dipolar in infinite systems\cite{Felder}  but 
extends  throughout the cell between metallic or polarizable  
 walls\cite{Takae}.  

We can   also study    propagation of sounds in glasses 
as a linear response to  a small-amplitude wall motion. In our 
theory,  its  time-evolution 
can be described  in terms of the   time-correlation functions 
of  the particle displacements  ${\bi u}_i(t)$ and ${\cal A}(0)$,  
where the  quasi-equilibrium  average is taken 
 around    a fixed inherent state. 
As in  granular materials\cite{Jia,Roux}, 
we shall find  rough wave fronts 
 and    random   scattered waves. 
It is of general interest how 
thermal sound waves come into play in   the 
 time-correlation functions 
in  films at low $T$.

The organization of this paper is as follows. In Sec. II, we
will  present the theoretical background  of the linear response 
in glasses  with respect to tangential 
 motions of the boundary walls.  In Sec.III, 
the linear response in supercooled  
and ordinary liquids will be briefly discussed. 
In Sec.IV, numerical results  will be 
presented to confirm our  theory in glasses. 
 Additionally,   a random elastic system 
will be treated in one dimension in Appendix A. 
  Correlations 
 among  the displacements and the wall  forces 
will be examined  for  homogeneous elastic moduli 
 in Appendix B.

\section{Linear Response at a fixed inherent state}

We consider a low-temperature glass 
composed of two species 
with  particle  numbers  $N_1$ and $N_2$. 
The total particle number is $N=N_1+N_2$. We write the particle 
positions  as 
 ${\bi r}_i=(x_i,y_i,z_i)$ and the  momenta  as 
${\bi p}_i =(p_i^x,p_i^y,p_i^z)$.
We assume nearly  rigid boundary walls at $z=\pm H/2$. 
These  particles are  confined  in the cell region $-H/2<z<H/2$  
along  the  $z$ axis, but   the periodic boundary condition is imposed  
 along the $x$ and $y$ axes with period $L$. The cell volume is 
  $V =HL^{d-1}$.    The lengths $H$ and $L$ 
are much longer than the particle diameters. 
Our   results can be used both 
in two and three dimensions ($d=2$ and $3$), 
where  the $y$ components are absent for $d=2$. 
In   Appendix A,   our theory will be presented 
in analytic forms in one dimension.

\subsection{Applying shear from boundary walls}

\begin{figure}
\includegraphics[width=0.9\linewidth]{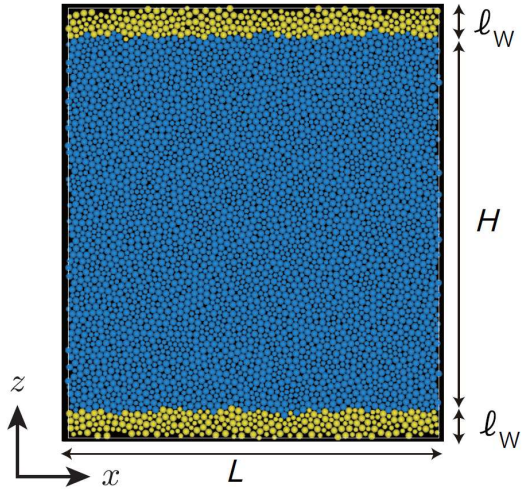}
\caption{(Color online) 
Illustration of geometry. Large and small 
particles (in blue) are in a glassy state   
 in the region $-H/2<z<H/2$. In 
top and bottom boundary layers
with thickness $\ell_{\rm w}$, particles (in yellow) 
in a glassy configuration 
are bound by rigid  springs to pinning centers ${\bi R}_k$ 
in the layers. 
The periodic boundary condition 
is imposed along the horizontal axes (axis  
for $d=2$) with period $L$. These unbound and bound 
particles interact via pairwise potentials. 
Their interfaces  are irregular and rigid, 
so  no slip occurs  when the 
walls are slightly shifted along  the $x$ axis.  
}
\end{figure}

As illustrated in Fig,1, we  induce    shear deformations
by motions of  boundary walls\cite{Shiba,Maeda}, which are  in the regions 
$-\ell_{\rm w}<z+H/2<0$ at the bottom 
and $0<z-H/2<\ell_{\rm w}$ at the top with $\ell_{\rm w}\ll H$. 
To each layer, $M$  particles are bound   
by spring  potentials $\psi (|{\bi r}_k -  {\bi R}_k|)$, where  
  ${\bi r}_k$ are the positions of these  particles and 
   ${\bi R}_k$  are the  pinning centers fixed to  the layers. 
We set   $N<k\le N+M$ at  the top  and  
$N+M<k\le N+2M$ at  the bottom. 
We assume the simple harmonic potential,  
\be 
\psi(r)=\frac{1}{2} s_0 r^2.
\en 
Pinning becomes stronger with increasing 
 the coefficient  $s_0$. 
The   bound particles also belong 
to either of the first or second species in the bulk 
and their  density is equal to the bulk density, so  $M= N \ell_{\rm w}/H$.

Particle pairs $i\in a$ and $j\in b$   (including the bound ones) 
interact  via short-ranged   potentials 
$\phi_{ab} (r_{ij}) $, where   $r_{ij}= |{\bi r}_i-{\bi r}_j|$ and 
 $a$ and $b$ denote the particle 
species ($1$ or $2$).  As a result, the  unbound particles do not penetrate into the boundary layers. For simplicity, we write $\phi_{ij}= 
\phi_{ab} (r_{ij}) $ and  $\psi_{k}= 
\psi (|{\bi r}_k -  {\bi R}_k|) $.  At    fixed ${\bi R}_k$,  
the  total potential energy  is given by   
\be 
U=  \frac{1}{2} \sum_{i,j} \phi_{ij}+ \sum_{k>N}  \psi_{k},
\en 
where we sum over all the particles in the first term 
and  the bound ones in the second  term $(k>N$).  
We write the   momentum density as   
${\bi J}= \sum_i  {\bi p}_i \delta ({\bi r}-
{\bi r}_i)$ using the $\delta$ function. 
Since the force  ${\bi f}_i= - \p U/\p {\bi r}_i$ 
on particle $i$ consists of the contributions 
 from the  particles and the walls, 
the  time derivative ${\dot {\bi J}}=\p {\bi J}/\p t$  
is written as 
\be 
{\dot {\bi J}}= -\nabla\cdot{\aP}-\sum_{k>N} \delta({\bi r}-{\bi r}_k)
\nabla_k \psi_{k}.
\en 
Here, $\aP=\{\Pi\a\}$ is the  microscopic  
  stress tensor\cite{Irving}  
 at position ${\bi r}$ and 
 the second term  is the force density from the walls.  
Hereafter, the over-dot denotes taking 
the time-derivative, 
$\alpha$ and $\beta$ represent 
 the Cartesian coordinates,  
and we set   $\nabla_i^\alpha= \p/\p x_i^\alpha$ 
and $\nabla_i=( \nabla_i^x, \nabla_i^y, \nabla_i^z)$.

We divide  the  stress tensor into  kinetic   and   potential parts 
as  $\Pi\a
= \Pi\a^{\rm K} +\Pi\a^{\rm p} $ with\cite{Irving,Onukibook}       
\bea 
\Pi\a^{\rm K}({\bi r}) &=& \sum_i \frac{1}{m_i} p_{i\alpha} p_{i\beta}  
\delta ({\bi r}-{\bi r}_i) ,\\   
\Pi\a^{\rm p} ({\bi r}) &=&  - \sum_{i,j}\frac{1}{2r_{ij} } \phi_{ij}'  
{x_{ij}^{\alpha} x_{ij}^{\beta}}\hat{\delta}({\bi r}, {\bi r}_i,{\bi r}_j) ,
\ena 
where    we sum over all the particles and set $\phi'_{ij}= 
d\phi_{ij}/dr_{ij}$  and $x_{ij}^{\alpha}= 
x_{i}^{\alpha}-x_{j}^{\alpha}$ (the $\alpha$ 
component of ${\bi r}_{ij}= {\bi r}_i-{\bi r}_j$). 
In Eq.(6), we  introduce   the  Irving-Kirkwood 
$\delta$ function  by 
\be 
{\hat \delta}({\bi r}, {\bi r}_i,{\bi r}_j)
= \int_0^1 d\lambda\delta ({\bi r}-\lambda {\bi r}_i -(1-\lambda){\bi r}_j) ,
\en 
which satisfies $\int d{\bi r}{\hat \delta}({\bi r}, {\bi r}_i,{\bi r}_j)
=1$ and  is nonvanishing only when $\bi r$ 
is on the line segment connecting ${\bi r}_i$ and ${\bi r}_j$. 
It follows   the relation  
${\bi r}_{ij}\cdot\nabla\hat{\delta}({\bi r}, {\bi r}_i,{\bi r}_j)= 
\delta ({\bi r}-{\bi r}_j)- \delta ({\bi r}-{\bi r}_i)$, 
 leading  to  $\sum_\beta \nabla_\beta  
\cdot{{\Pi}}^{\rm p}_\ab =\sum_{i,j}
 \delta ({\bi r}-{\bi r}_i)\nabla_i^\alpha \phi_{ij}$ in 
Eq.(4). If we  integrate $z  {\dot {J}}_\alpha$ 
in a region  containing  all the particles,  
Eq.(4) gives  a useful relation, 
\be 
\int d{\bi r}\Pi_{z\alpha}({\bi r}) 
-  \int d{\bi r} z {\dot { J}}_\alpha ({\bi r}) 
=   \sum_{k>N}
z_k \nabla_k^\alpha \psi_k,
\en 
where the right hand side arises from the wall potentials.

We next  move the top wall by $\gamma_{\rm ex} H/2$ 
and the bottom  wall by $-\gamma_{\rm ex} H/2$ along the $x$ axis. 
Here  $\gamma_{\rm ex}$ 
is a small mean  shear strain, which can depend on time. 
Then, in $U$ in  Eq.(3),  the positions  
 ${\bi R}_k$ are shifted by $\pm \gamma_{\rm ex} H/2$ 
along the $x$ axis and  $\psi_k$  
are changed by $\mp (\gamma_{\rm ex} H/2)(\p \psi_k /\p x_k)$ 
 to  linear order in $\gamma_{\rm ex}$. 
Thus,  $U$ is changed 
by  $ -\gamma_{\rm ex} {\cal A}$ with  
\be
{\cal A}=\frac{H}{2}
(F_{\rm bot}^x- F_{\rm top}^x)= \frac{H}{2} \Big(
\sum_{k\in {\rm top} } -
\sum_{ k\in {\rm bot} }\Big)\nabla_k^x\psi_k .
\en  
where   $F_{\rm top}^\alpha= -  \sum_{ k\in {\rm top}}\nabla_k^\alpha 
\psi_k$ is  the  force on the particles   from the top wall 
 and  $ F_{\rm bot}^\alpha=-\sum_{ k\in {\rm bot}}\nabla_k^\alpha  
\psi_k$ is that from the bottom wall. 
In Eq.(9), $\cal A$ consists of   
the contributions  from  the bound particles ($k>N$) and  
 is    amplified by the   prefactor $H/2$. 
Hereafter,   for any $a_k$, we write $ (\sum_{k\in {\rm top} } -
\sum_{ k\in {\rm bot} }) a_k= 
\sum_{k\in {\rm top}} a_k  -\sum_{k\in {\rm bot}}a_k$. 
Previously,  Shiba and one of the present authors\cite{Shiba} 
applied  a shear flow  in the geometry in Fig.1 
 to binary particle systems in  various  states.

For $\alpha=x$, the right hand side of Eq.(8) 
is rewritten as ${\cal A}+   \sum_{k>N}
\delta z_k \nabla_k^x \psi_k$, where 
 $\delta z_k= z_k \mp H/2$  at the top and  the bottom layers, so 
 it is nearly equal to $\cal A$ 
 for  $|\delta z_k|< \ell_{\rm w}\ll H$.
Thus,  $\cal A$  can also be    expressed  as  
a  difference of two bulk integrals, 
\bea 
{\cal A}&=& \int d{\bi r}\Pi_{zx}({\bi r})-\int d{\bi r} z{\dot{J}}_x({\bi r}) 
\nonumber\\
&=& \int d{\bi r}\Pi_{zx}^{\rm p}({\bi r})- \sum_i z_i f_i^x.  
\ena  
Here, the kinetic parts cancel  from  the relation  
  $\int d{\bi r} z  {\dot J}_z=
 d( \sum_i z_i p_i^x)/dt= 
 \sum_i[p_i^zp_i^x/m_i + z_i f_i^x]$.
In the second line,   $f_i^x$ is the $x$ component 
of the force ${\bi f}_i=- \p U/\p {\bi r}_i$ 
on particle $i$, so 
the sum $\sum_i z_i f_i^x$ is the off-diagonal virial. 
Since   $U $ includes 
the  wall potentials,  the total force on the particles 
is given by  
\be 
\sum_i {\bi f}_i= {\bi F}_{\rm top}+{\bi  F}_{\rm bot}.
\en

Note that we can also  move the top wall  by $\gamma_{\rm ex}(H/2-a)$ 
and the bottom wall  by $-\gamma_{\rm ex}(H/2+a)$, where   
$a$ is an arbitrary length. 
 For example, the top  wall is at rest   for   $a=H/2$. 
 Then, ${\cal A}$ in Eq.(9) is changed to
 ${\cal A}'= (H/2+ a)F_{\rm bot}^x - (H/2-a) F_{\rm top}^x$ and  
 $z_i$ in  Eq.(10) is changed to $z_i- a$. 
  Our  expressions for   $\mu$ 
will not depend on  $a$ (see the sentences below Eqs.(44) and (50) 
and Eq.(51)).

\subsection{Inherent state and quasi-equilibrium in glass }

Glasses  are nonergodic at low $T$, where   
 the configuration changes are negligible during the observation 
 in the limit of small  applied strain. 
In the corresponding inherent state in the limit $T\to 0$, 
we write the particle  positions    as  ${\bi r}_i = {\bi r}_i^\ih= 
(x^\ih_i,y^\ih_i,z^\ih_i)$, for which  
the mechanical equilibrium (${f}_i^\alpha =-\nabla_i^\alpha  {U}=
{ 0}$) holds for  all the particles. Thus, each inherent state is 
one of the local minima in  the many-particle phase space. 
In addition,    the inherent (residual) shear stress 
  $\Pi_\ab^{\rm ih}({\bi r})
=\lim_{T \to 0} \Pi_\ab({\bi r})$ 
 is   highly heterogeneous 
with  long-range correlations\cite{Harro,Egami,Lema,Fuchs,Seme}.  
In our case, from Eqs.(9) and (10), the total 
shear stress is related to the  force difference  as   
\be 
W^\ih=
 \int d{\bi r}\Pi_{zx}^{\rm ih}({\bi r})= 
\frac{1}{2}H(F_{\rm bot}^x- F_{\rm top}^x)
\quad (T\to 0), 
\en 
while the total applied 
force  ${\bi F}_{\rm bot}+{\bi F}_{\rm top}$   vanishes 
 as $T\to 0$ from Eq.(11). 
Fuereder and Ilg\cite{Ilg} found a counterpart of Eq.(12)  
in the periodic boundary condition. 
See more discussions on $\Pi_{zx}^{\rm ih}({\bi r})$ in Sec.IIH.

We  next consider  the  particle displacements, 
\be 
{\bi u}_i= (u_i^x,u_i^y,u_i^z)
={\bi r}_i- {\bi r}_i^{\ih},
\en 
 from   the inherent positions.     
To leading order,   the effective Hamiltoninan for ${\bi u}_i$ 
  is of  the  bilinear form,
\be    
{\cal H}_\ih=  \sum_{i} \frac{1}{2m_i} |{\bi p}_{i}|^2 + 
\delta U,
\en 
where  ${\bi p}_i= m_i {\dot {\bi u}}_i$  and  the potential 
part $\delta U$ is given by  
\be 
\delta U =\frac{1}{4}  \sum_{i,j,\alpha,\beta}\hspace{-1mm} 
{s_{ij}^{\ab} } { u}_{ij}^\alpha u_{ij}^\beta  + 
\frac{1}{2} 
\hspace{-0.5mm}\sum_{k>N,\alpha}  \hspace{-1.2mm}
s_{0} |{{u}_{k}^\alpha }|^2.
\en 
The first term depends on 
$u_{ij}^\alpha= u_i^\alpha-u_j^\alpha$.
We define 
\be
 s_{ij}^{\ab}= \nabla_i^\alpha \nabla_i^\beta
\phi_{ij}=  (\frac{\phi_{ij}''}{r_{ij}^2} 
-\frac{\phi_{ij}'}{r_{ij}^3}) x_{ij}^\alpha x_{ij}^\beta+  
\frac{\phi_{ij}'}{r_{ij}}\delta_\ab,
\en 
where   $\phi_{ij}''= d^2\phi_{ij}/dr_{ij}^2$ 
and the particle  positions    are  at 
${\bi r}_i = {\bi r}_i^\ih$.    For simplicity, 
 $x_i^\ih$, $y_i^\ih$, and $z_i^\ih$ will be 
written as  $x_i$, $y_i$, and $z_i$,  
when    confusion will  not  occur.

The    ${\cal H}_\ih$  describes  the local 
 vibrational   motions  
in local potential minima   
and the collective acoustic modes on larger scales. 
We assume that the the observation time is much longer than 
the microscopic  times but  is much shorter 
than the structural relaxation time (see Sec.IIIA).  
Then,  quasi-equilibrium should be attained 
at fixed ${\bi r}_i^\ih$, 
where  $({\bi u}_i,{\bi p}_i$) obey  the   canonical distribution,
\be 
P_{\rm ih}\propto \exp (- {\cal H}_\ih /k_BT).
\en 
Hereafter, the thermal average over this  distribution 
 will be written as 
$\av{\cdots}_\ih= \int \Pi_{i=1}^{N+2M} d{\bi p}_i d{\bi u}_i 
(\cdots )P_\ih$ for a fixed  inherent state ({\it isoinherent ensemble}). 
For simplicity, we assume the Gaussian  
form of $P_\ih$ to  obtain 
 $\av{u_i^\alpha u_j^\beta}_\ih \propto k_BT$. However, 
 there is no difficulty  to include the  anharmonic 
potential terms, which can be important with increasng $T$ 
(see a remark  below Eq.(42))\cite{Sastry}. 
  

To linear order, the force on particle 
$i$ is written as
\be 
f_i^\alpha= -\frac{ \p}{\p u_i^\alpha} {\delta U} 
= -\sum_{j,\beta}
h_{ij}^\ab u_j^\beta. 
\en 
From Eq.(15) the Hessian    matrix   $\{h_{ij}^\ab\}$ is 
symmetric as  
\be 
h_{ij}^\ab=\sum_k s^\ab_{ik}\delta_{ij}  -s_{ij}^\ab + 
 s_0 \delta_\ab\delta_{ij}\theta_{i-N},
\en 
where   $\theta_{i-N}$ is  1 for $i> N$ and  0 for $i\le N$. 
The  last term is the 
 contribution from  the bound particles 
at the  walls and is nonexistent in  the periodic 
boundary condition. 

The potential energy deviation in Eq.(15)  
 assumes  the symmetric bilinear form 
$\delta U= 
\sum_{i,j,\alpha,\beta} h_{ij}^\ab u_i^\alpha u_j^\beta/2$. Thus,  
the inverse matrix of   $\{h_{ij}^\ab\}$ 
is related to the displacement variances as 
 \be 
 (h^{-1})_{ij}^\ab= \av{u_i^\alpha u_j^\beta}_\ih/k_BT, 
\en 
From   $f_i^\alpha= k_BT \p(\ln P_\ih)/\p u_i^\alpha$,  
any variable $\cal B$ satisfies 
\be   
-\av{{\cal B}{f}_i^\alpha}_\ih=\sum_{k,\beta} 
 h_{ik}^{\alpha\beta} 
\av{ u_k^\beta {\cal B}}_\ih= 
 k_BT \av{{\p{\cal B}}/{\p u_i^\alpha}}_\ih.
\en 
For ${\cal B}= u_i^\alpha$  we  find   
$
 \av{u_i^\alpha f_j^\beta}_\ih=-k_BT \delta_{ij}\delta_\ab. 
$   See a similar relation in Eq.(69) in equilibrium.


Many authors\cite{Sch,Liu,Yama,Gelin,Barr1,Barr2,Malo,Elliott} 
have  calculated   the vibrational  modes in glasses 
  in  the periodic boundary 
condition.  We can  examine them with  boundary walls 
by including the last term in  Eq.(19)  
(see Fig.2). 

\subsection{ $\Pi_\ab^{\rm p}$ and $\cal A$  
around an inherent state} 

Around an inherent state, the potential part of the stress tensor
 $\Pi_\ab$ in Eq.(6)  is  expanded as
\be   
\Pi^{\rm p}_\ab ({\bi r}) = \Pi_\ab^\ih({\bi r})  
+ \delta\Pi^{\rm p}_\ab({\bi r}) +\cdots.
\en 
The first term is the  inherent stress  
and the second term is the first-order deviation linear in ${\bi u}_i$ given by \bea
&&\hspace{-1cm}  
\delta\Pi^{\rm p}_\ab({\bi r}) = - \frac{1}{2} \sum_{i,j,\nu}
 w_{ij}^{\ab\nu} u_{ij}^\nu {\hat\delta}({\bi r}, {\bi r}_i,{\bi r}_j)
\nonumber\\
&&\hspace{-5mm} +  \sum_{i,j} 
\frac{{\phi_{ij}'}{x_{ij}^\alpha}}{r_{ij}} ({\bi u}_i\cdot\nabla)  
[(x_\beta -x_j^\beta)  {\hat\delta}({\bi r}, {\bi r}_i,{\bi r}_j) ].  
\ena 
In the first term the  coefficients $w_{ij}^{\ab\nu}(=w_{ij}^{\beta\alpha\nu}
=-w_{ji}^{\ab\nu}$) are defined in the inherent state by
\be
 w_{ij}^{\ab\nu} = \nabla_i^\nu  [ \phi_{ij}'  
{x_{ij}^{\alpha} x_{ij}^{\beta}}/r_{ij}]
=  s_{ij}^{\alpha\nu}  x_{ij}^\beta + 
\nabla_i^\alpha\phi_{ij} \delta_{\beta\nu} . 
\en 
We derive the second term in Eq.(23)  from 
the deviation of $\hat\delta$ in Eq.(7) using  
  $x_{ij}^\beta {\bi u}_i\cdot \nabla_i  {\hat\delta}
=- {\bi u}_i\cdot \nabla [(x_\beta- x_j^\beta) {\hat\delta}]$, which 
 vanishes upon  space integration. In this paper, we consider  
the space integral of $\delta\Pi_{zx}^{\rm p}({\bi r})$ written as  
\bea 
 \delta W_{\rm p}&=& 
 \int d{\bi r}\delta\Pi_{zx}^{\rm p}({\bi r})
= -  \frac{1}{2}\sum_{i,j,\alpha}
 w_{ij}^{zx\alpha} {u_{ij}^\alpha} 
\nonumber\\
&&
= -\sum_{i,\alpha}
w_{i}^{\alpha}u_{i}^\alpha, 
\ena  
where we can replace $u_{ij}^\nu= u_i^\nu-u_j^\nu$ 
in the first line by $2u_i^\nu$ 
to obtain the second line and   we define the coefficients, 
\be
 w_i^\alpha
= \sum_j w_{ij}^{zx\alpha}
= \sum_j[ s_{ij}^{z\alpha}x_{ij} + 
\frac{\phi_{ij}'}{r_{ij}}z_{ij}\delta_{x\alpha}].
\en  

From  Eqs.(9), (10), and (25)  
the  deviation of  $\cal A$  is written to 
linear order  in the following two forms,        
\bea 
&&\delta {\cal A}= \frac{1}{2}H (\delta F_{\rm bot}^x-
\delta F_{\rm top}^x)
\nonumber\\
&&\hspace{5mm}
 = \delta W_{\rm p} - \sum_i z_i f_i^x. 
\ena
Here,  $\delta F_{\rm top}^\alpha$ is  the deviation of 
the  force   from the top wall 
and $\delta F_{\rm bot}^\alpha$ is that from the bottom 
wall, so    
\be 
\delta F_{\rm top}^\alpha = -s_0 
\sum_{ k\in {\rm top}}u_k^\alpha, \quad 
\delta F_{\rm bot }^\alpha = -s_0 
  \sum_{k\in {\rm bot}} u_k^\alpha. 
\en 
We can  also write     
$\delta{\cal A}=  s_0 \sum_{k>N} z_k u_k^x$ 
for  $H\gg \ell_{\rm w}$.


\subsection{Linear response in   quasi-equilibrium}

For   a fixed   inherent state, we use Kubo's method\cite{Kubo} in 
 the linear response theory for the perturbed Hamiltonian,
\be 
{\cal H}'_\ih= \sum_i \frac{1}{2m_i}|{\bi p}_i|^2  +\delta U  - \gamma_{\rm ex}
(t) \delta{\cal A}
\en  
around the quasi-equilibrium distribution 
 $P_\ih$  in Eq.(17),  
where   ${\delta U} $ 
and ${\delta {\cal A}} $ are given by Eqs.(15) and (27). 
The equations of motion are  $\dot{\bi p}_i= {\bi f}_i 
+ \gamma_{\rm ex}\p ( \delta{\cal A})/\p {\bi u}_i$ 
including  the perturbation. 
As a result, the phase-space 
distribution  slightly deviates from $P_\ih$. 
For any variable ${\cal B}$,  we consider its time-dependent average 
over  this perturbed distribution.  
Its deviation  due to  $\gamma_{\rm ex}$ is given  by   
\be 
\delta {\bar{\cal B}}(t) = \chi_{BA}(0) \gamma_{\rm ex} (t) - 
\int_{0}^t  ds \chi_{BA}(t-s) {\dot\gamma}_{\rm ex}(s) , 
\en 
where ${\dot \gamma}_{\rm ex}(t)= d\gamma_{\rm ex}(t)/dt$ 
and we assume  ${ \gamma}_{\rm ex}(t)=0$ for $t<0$. 
We define the  response   function, 
\be 
 \chi_{BA}(t)= \av{{\cal B}(t) \delta{\cal A}(0)}_\ih/k_BT ,
\en 
where the  time-evolution  is governed by the unperturbed 
Hamiltonian  ${\cal H}_\ih$ in Eq.(14) (see Sec.IIG). 
The  static susceptibility is  given by 
 the equal-time correlation in Eq.(30). 
Use of  the second line of Eq.(27) gives    
\be 
\chi_{BA}(0)= \frac{\av{{\cal B}{\delta\cal A}}_\ih}{k_BT}= 
 \av{\sum_i  z_i \frac{\p{\cal B}}{\p u_i^x}}_\ih
+ \frac{ \av{{\cal B}  \delta W_{\rm p}}_\ih}{k_BT},
\en 
where the  first term arises from the affine change  
in $\cal B$ and   the second term is due to the  
correlation between $\cal B$ and  the 
deviation $\delta W_{\rm p}$ in the total shear stress.

In particular, for a stepwise shear strain $\gamma_{\rm ex}(t)
= \gamma_0 \theta(t)$  
(being  0 for $t<0$ and  $\gamma_0  $ for $t>+0$), 
we find 
\be 
\delta {\bar{\cal B}}(t) /\gamma_0 =
 \chi_{BA}(0) - \chi_{BA}(t) \quad (t>0).
\en   
For   ${\cal B}= \delta W_{\rm p}$, we  
obtain the stress relaxation function $G_\ih (t)= 
 -\delta {\bar{W}}_{\rm p}(t) /\gamma_0$ for $t>0$. 
From Eq.(33)  we find     
\be
G_\ih (t) =\mu+ \av{\delta W_{\rm p}(t) \delta{\cal A}(0)}_\ih/k_BT V , 
\en 
where $\mu$ is the average shear 
modulus in Eq.(38) below, so  $G_\ih(0)=0$.  
Our $\chi_{BA}(t)$ 
and $G_\ih(t)$ exhibit oscillatory 
behavior and can be used only for   $t$  
before appreciable structural relaxation (see Fig.7). 
However, in the  literature of rheology\cite{Saw,Doi,Wit1}, 
the stress relaxation function   
is the sum of $\mu$ and  the stress time-correlation function 
in  Eq.(61) (divided by $k_BT$).

\subsection{Displacement and stress  in  steady shear strain }

Let us make further calculations for   a  steady shear strain 
$\gamma_{\rm ex}$. For 
${\cal B}= u_i^\alpha$,  the first line of Eq.(27)  
  gives   the average dispacements. With the aid of Eq.(28) we find   
\be 
\frac{{\bar u}_i^\alpha}{\gamma_{\rm ex}}  
= \frac{\av{{u_i^\alpha}{\delta\cal A}}_\ih}{k_BT}
= \frac{Hs_0}{2}
\Big (
\sum_{k\in {\rm top} } -
\sum_{ k\in {\rm bot} }\Big)  
 (h^{-1})_{ik}^{\alpha x}.
\en 
Here,   all $u_i^\alpha$ are coupled with 
the   force difference 
$\delta F_{\rm bot}^x- 
\delta F_{\rm top}^x $. See   Sec.IIF  for   more discussions 
on  $\delta F_{\rm bot}^x$ and $\delta F_{\rm top}^x $. 
Furthermore,  Eq.(32) gives another expression,  
\be 
\frac{{\bar u}_i^\alpha}{\gamma_{\rm ex}}
  = z_i \delta _{\alpha x}
-  \sum_{k,\beta} (h^{-1})_{ik}^{\alpha\beta} w_{k}^{\beta},   
\en 
which   consists    of  the affine and nonaffine parts\cite{Malo}.

For   ${\cal B}= \delta\Pi_{zx}^{\rm p}({\bi r})$, we   define the  local 
shear modulus by 
\be 
{\hat \mu}({\bi r})=-\av{\delta{{\bar \Pi}}_{zx}^{\rm p}({\bi r})}
/\gamma_{\rm ex}= 
 -\av{{\delta{\Pi}_{zx}^{\rm p}({\bi r})}\delta{\cal A}}_\ih/k_BT .
\en
This ${\hat \mu}({\bi r})$  has  the particle 
discreteness in an inherent state, so we need to 
integrate it in  small squares or cubes in  the cell  
to detect  elastic heterogeneity\cite{Miz1,PabloP}.
The   space average of  ${\hat \mu}({\bi r})$ in the whole cell 
is simpler as    
\be 
\mu = \int d{\bi r} 
\frac{{\hat{\mu}} ({\bi r})}{V}=-
\frac{\av{\delta W_{\rm p}\delta{\cal A}}_\ih}{Vk_BT}=  
\sum_{i,\alpha} \frac{ w_{i}^{\alpha} {\bar u}_{i}^\alpha 
}{V\gamma_{\rm ex} }, 
\en 
Then, as $T\to 0$, Eqs.(35) and (38) yield  
\be 
{\mu} =
 \frac{s_0}{2V} H\sum_{i,\alpha} \Big (
\sum_{k\in {\rm top} } -
\sum_{ k\in {\rm bot} }\Big) w_i^\alpha (h^{-1})_{ik}^{\alpha x} . 
\en 
which arises from the correlations  between the bound and unbound particles.
 
From Eq.(36) we also find the well-known expression,  
\be
{\mu} =\av{ \mu_\infty}_\ih   
- {{\av{ \delta W_{\rm p}^2}_\ih}}/{Vk_BT} . 
\en 
where $\mu_\infty= \sum_i  w_i^x z_i/V$ is the affine 
contribution,  
\be 
\mu_\infty= 
 \frac{1}{2V} \sum_{i,j}\bigg[ 
 (\frac{\phi_{ij}''}{r_{ij}^2} 
-\frac{\phi_{ij}'}{r_{ij}^3})  
x_{ij}^2 {z_{ij}^2}  +\frac{\phi_{ij}'}{r_{ij}}{z_{ij}^2} \bigg]   
\en 
This   $\mu_\infty$ was first introduced 
for fluids (see  Eq.(57))\cite{Zwanzig}. 
As $T\to 0$, $\av{ \mu_\infty}_\ih $ is equal to 
$ \mu_\infty$ at the inherent positions ${\bi r}_i= {\bi r}_i^\ih$, while 
  the    nonaffine part is proportional to the 
stress variance and is negative. As $T\to 0$,  we find      
\be
\av{ (\delta W_{\rm p})^2}_\ih /Vk_BT 
= \sum_{i,j,\alpha,\beta}
(h^{-1})^\ab_{i,j}w_i^\alpha w_j^\beta /V. 
\en 
We can use  Eq.(40)  even if we include the anharmonic potential 
terms in the average $\av{\cdots}_\ih$ (see a comment below Eq.(17)). 
For  large $V$,  the bulk contributions  
dominate over the surface ones  in  Eqs.(40)-(42).
See Appendix A for counterparts of Eqs.(40)-(42) in one dimension.

From Eqs.(27) and (30) the average of $\delta{\cal A}$ is given by   
\be
{\delta{\bar{\cal A}}}/{\gamma_{\rm ex}}
={\av{(\delta{\cal A})^2}_\ih}/{k_BT}= s_0 H^2M/2-\mu V, 
\en
where we use  
 $\av{\delta{\cal A}\sum_i z_if_i^x}_\ih/k_BT=  
 - s_0\sum_{i>N}  z_i^2$. 
Since $\av{\delta{\cal A}^2}_\ih>0$, 
 we  require 
$s_0 > 2\mu V/H^2M$, 
which is well satisfied in our simulation in Sec.IV. 
From  Eqs.(27), (28),  and (43) we 
can also  express $\mu$ as 
\be 
\mu= \frac{ 1}{V} \sum_{i>N,j>N} 
{z_i z_j }[s_0 \delta_{ij}- s_0^2  
\av{u_i^xu_j^x}_\ih/k_BT ] .
\en 
in terms of  the variances of the  bound particles. 
This relation  is invariant with respect to 
the coordinate shift along  the $z$ axis 
($z_i \to z_i -a$ and $z_j \to z_j -a$) (which can be 
proved  from  Eqs.(47) and (48) below).

In Eq.(15) we  replace 
    $u_k^x$  by $u_k^x \mp  H\gamma_{\rm ex}/2$ 
for $k>N$ to obtain the change in  
 the  potential energy  deviation,    
\be
\delta U'=\delta U 
 - \gamma_{\rm ex}\delta {\cal A}+
\gamma_{\rm ex}^2 s_0 H^2M/2, 
\en 
up to  of order $\gamma_{\rm ex}^2$, 
 Minimization of  $\delta U'$  with respect to $u_i^\alpha$ 
gives $ f_i^\alpha+ \gamma_{\rm ex} 
\p (\delta {\cal A})/\p u_i^\alpha=0$. 
This leads  to  Eqs.(35) and (36) with the aid of 
the first and second lines of  Eq.(27). 
Note that this derivation of Eq.(36) 
is equivalent to that of 
  Maloney and Lema\^itre\cite{Malo,Malo1}.
Let us  then calculate  the minimum value 
of $\delta U'$ by seting   $u_i^\alpha$  
equal to $ {\bar u}_i^\alpha$ in Eq.(35) 
and  $\delta {\cal A}$  equal to 
 $\delta{\bar{\cal A}}$ in Eq.(43).  
In terms of $\mu$ in Eq.(38), it is simply  written as         
\be 
\delta {\bar U}'=  \mu \gamma_{\rm ex}^2V/2,
\en 
which  is not  obvious for inhomogeneous 
 glassy systems. 

We note the following. 
(i) For crystals,  the shear moduli  are 
expressed  in  the same form as  in  Eq.(40), 
where the thermal average is  taken 
over the displacement fluctuations  in  a given crystal 
state\cite{Hess,Ray,Lutsko,Hoover}. 
For  glasses, the average is taken 
over   one  inherent states\cite{Malo,Malo1,Evans,Yoshi,Ilg,Saw}. 
(ii) To examine elastic  inhomogeneity  
in glasses, some authors\cite{Miz1,PabloP}
  divided   the cell into small regions 
and integrated  the average of the first term in Eq.(23) 
in each region (where $\hat{\delta}$ can be integrated in a simple form).

\subsection{Fluctuations of forces from walls}

We have introduced the forces from the walls to the particles 
in Eq.(9).   We here examine  the equal-time correlations 
among  their thermal fluctuations 
 and $u_i^\alpha$. From Eqs.(11) and (21) and 
$\sum_j h_{ij}^\ab= -s_0 \theta_{i-N}\delta_\ab$,  
 the total force deviation     $\delta F_{\rm tot}^\alpha
\equiv  \delta F_{\rm top}^\alpha+ 
\delta F_{\rm bot}^\alpha$ satisfies      
\bea 
&&\hspace{-1.2cm} \av{u_i^\alpha 
\delta F_{\rm tot}^\beta}_\ih=-  k_BT   \delta_\ab ,\\
&&\hspace{-12mm} 
\av{ \delta F_{\rm tot}^\alpha \delta F_{\rm tot}^\beta}_\ih  
= 2k_BT s_0 M\delta_{\alpha\beta} . 
\ena  
From Eq.(11)  the left hand side of  Eq.(47)  
is written  as  $\av{u_i^\alpha \sum_j {\dot p}_j^\beta}_\ih  
=- \av{{\dot u}_i^\alpha  \sum_j p_j^\beta}_\ih$; 
then, Eq.(47) is obvious.   From Eqs.(25) and (26)
$\delta W_{\rm p} $ and $\delta F_{\rm tot}^\alpha$ are orthogonal as 
\be 
\av{\delta W_{\rm p} \delta F_{\rm tot}^\alpha}_\ih =0. 
\en 

On the other hand, the  $x$-component of the  force difference  
$\delta F_{\rm top}^x- \delta F_{\rm bot}^x$  is proportional 
 to $\delta{\cal A}$ as in Eq.(27), so its 
relations follow from  Eqs.(35) and (43).   
Using  Eq.(48) also, we find the cross correlation, 
\be  
\av{ \delta F_{\rm top}^x \delta F_{\rm bot}^x}_\ih /k_BT= \mu V/H^2=
\mu L^{d-1}/H.  
\en  
 See   Appendices  A and  B for  counterparts 
of Eq.(50) in  simpler situations.
Here, we  argue that Eq.(50) is  general for elastic films. 
Let us move the bottom layer by $-H \gamma_{\rm ex}$ 
with the top layer kept at rest (see  the last paragraph of Sec.IIA). 
 Then, $\delta {\cal A}$ is changed to $\delta {\cal A}'= 
H \delta F_{\rm bot}^x$ in Eq.(29)  and 
the average  force from  the top wall 
to the particles   is given by 
$\av{ \delta F_{\rm top}^x \delta{\cal A}'}_\ih /k_BT= \mu V/H$ 
in equilibrium, which coincides with  Eq.(50). 
 See Eq.(82) and Fig.8  for the 
 time-correlation of the wall forces. In accord with this 
argument, Eqs.(38) and (49) give    
\be 
\mu
=- {\av{\delta W_{\rm p}\delta{F}_{\rm bot}^x) }_\ih}H/{Vk_BT},
\en
where $\delta{F}_{\rm bot}^x$ can be replaced by $-\delta{F}_{\rm top}^x$.

\subsection{Linear dynamics around  a fixed inherent state }
  
Around  a fixed inherent state, 
 the  dynamic equations follow from  ${\cal H}_\ih$ in Eq.(14). They are 
 rewritten  as 
\be 
\frac{d^2}{dt^2}  {\hat  u}_i^\alpha(t) = - 
\sum_{j,\beta}{\hat h}_{ij}^\ab {\hat u}_j^\beta(t).
\en 
 For $m_1\neq m_2$,       we use the 
following scale changes\cite{Elliott,Yama},  
\be 
{\hat u}_i^\alpha(t)= \sqrt{m_i} { u}_i^\alpha(t), 
\quad 
{\hat h}_{ij}^\ab=  { h}_{ij}^\ab/\sqrt{m_i m_j}, 
\en 
where ${\hat h}_{ij}^\ab$  is 
 the modified Hessian   matrix. Its inverse 
 is given by 
$
{ ({\hat h}^{-1})}_{ij}^\ab= 
\sqrt{m_im_j} { (h^{-1})}_{ij}^\ab$.   
 We  introduce the 
$d(N+2M)$ dimensional 
eigenvectors $e^\lambda_{i\alpha}$ satisfying 
\be 
\sum_{j,\beta} {\hat h}_{ij}^{\ab} e^\lambda_{j\beta }
= \omega_\lambda^2 e^\lambda_{i\alpha} ,
\en 
where     $0<\omega_1\le \omega_2\le \cdots$. 
The $e_{i\alpha}^\lambda$ are  normalized as  
$\sum_\lambda e^\lambda_{i\alpha}e^\lambda_{ j\beta}=
 \delta_\ab\delta_{ij}$ and  
$\sum_{i,\alpha} e^\lambda_{i\alpha}e^\sigma_{ i\alpha}=
 \delta_{\lambda \sigma}$.  
Projection of  $ {\hat u}_i^\alpha$ 
on the eigenmodes yields 
 the   variables $s_\lambda$ as   
\be 
 s_\lambda= \sum_{i,\alpha} e^\lambda_{i\alpha} {\hat u}_i^\alpha, \quad 
{\hat u}_i^\alpha= \sum_\lambda e_{i\alpha}^\lambda s_\lambda. 
\en 
The potential energy deviation in Eq.(15) is  
written  as 
\be 
\delta U= \frac{1}{2}\sum_{i,\alpha,j,\beta}{\hat h}_{ij}^{\ab}
{\hat u}_i^\alpha{\hat u}_j^\beta
=
\frac{1}{2}\sum_\lambda \omega_\lambda^2 s_\lambda^2,
\en
from which we find  
$\av{ s_\lambda s_\sigma}_\ih= k_BT  
\delta_{\lambda\sigma} /\omega_\lambda^2$. 

Now, from Eqs.(52)-(55), we  obtain 
the dynamic equations    
$d^2 s_\lambda (t)/dt^2 = -\omega_\lambda^2 
s_\lambda(t)$.  These are  solved to give 
$
s_\lambda(t) = s_\lambda(0) \cos(\omega_\lambda t)+
{\dot s}_\lambda(0) \sin(\omega_\lambda t)/\omega_\lambda.
$ with  ${\dot s}_\lambda(0)= 
 \sum_{i,\alpha} e^\lambda_{i\alpha}\sqrt{m_i} {\dot  u}_i^\alpha(0)$ 
being  linear in  the  velocities at $t=0$. Thus,   averaging  over $P_\ih$ 
yields   
\be 
\av{ s_\lambda(t) 
 s_\sigma(0)}_\ih/k_BT  =
\delta_{\lambda\sigma} 
 \cos(\omega_\lambda t) /\omega_\lambda^2.
\en 
For any  variable  $\delta{\cal B}(t)= 
\sum_{i,\alpha}b_i^\alpha u_i^\alpha(t)$ 
linear in $u_i^\alpha(t)$, we can express it as 
$\delta{\cal B}(t) = \sum_\lambda Z_B^\lambda s_\lambda(t)$, where 
\be 
Z_B^\lambda= \sum_{i,\alpha} {b_i^\alpha}
 e_{i\alpha}^\lambda/{\sqrt{m_i}}.
\en 
Furthermore, if $\delta{\cal A}(t) = \sum_\lambda Z_A^\lambda s_\lambda(t)$, 
Eq.(57) gives  
\be 
 \chi_{BA}(t)= 
\frac{\av{\delta{\cal B}(t)\delta{\cal A}(0)}_\ih }{k_BT} =
\sum_\lambda  Z_B^\lambda Z_A^\lambda
 \frac{ \cos(\omega_\lambda t)}{\omega_\lambda^2}. 
\en 
At $t=0$,  Eq.(59) is the equal-time 
correlation from 
$\sum_\lambda e^\lambda_{i\alpha}\omega_\lambda^{-2}
e^\lambda_{j\beta}={ ({\hat h}^{-1})}_{ij}^\ab$.
The expressions (57) and  (59) exhibit   oscillation  even at long 
times without dissipation\cite{Gelin,Elliott}. 
In particular, if $\delta{\cal B}= u_{i}^\alpha$ 
and $\delta{\cal A}$ is given by Eq.(27),  
 Eqs.(35), (55), and (59) lead to  another expression for the 
average induced displacements, 
\be 
{\bar u}_i^\alpha/\gamma_{\rm ex} = \sum_\lambda  e_{i\alpha}^\lambda Z_A^\lambda/(\sqrt{m_i}\omega_\lambda^2). 
\en 

\subsection{Stress time-correlation in simulation}

The stress time-correlation function    has been   calculated via 
 molecular dynamics simulation\cite{Onukibook,Harro,Zwan,Hansen,Green}.  
 Using   the integral of the  shear stress $W(t) = \int d{\bi r} 
\Pi_{zx}({\bi r},t)$ at time $t$ for large systems, we express  it as  
\be 
C(t)=    \av{W(t+t_0) W(t_0)}_{\rm sim}/V.  
\en    
As in  usual simulations,  $\av{\cdots}_{\rm sim}$ includes    
the  average     over a long time interval of the initial time 
$t_0 \in  [0, t_{\rm sim}]$. It  can also 
be over many   simulation runs or over many   inherent states in glasses. 
For a suitable  ensemble, the integral of the inherent stress 
$W^\ih=\int d{\bi r} \Pi^\ih$ can obey a distribution with   
$ \av{ W^\ih}_{\rm sim}=0$  without  applied strain\cite{Harro,Ilg}. 
 Here, if $t$ and $t_{\rm sim}$ are sufficiently long    
at not very small $T$,  
  $C(t)$ slowly decays  to zero due to the configuration changes 
(see Sec,IIIA).

It is well-known that   $C(t)$  decays from its initial value 
$C(0)$ to a well-defined plateau value 
$ C_{\rm pl}$ after a microscopic time at low $T$. 
In our simulation in Sec.IV, 
this will be the case after  averaging over $t_0$  even 
in  a single  run. 
Here, $C(0)$ and $C_{\rm pl}$ do not depend 
on  the system size  for large $V$. 
The  $ C_{\rm pl}$ is nearly independent of   $T$ and is  related to 
 the inherent stress as\cite{Ilg,Harro,Yoshi,Evans,Wil} 
\be
 C_{\rm pl}= \av{ (W^\ih)^2}_{\rm sim}/V \quad (T\to 0), 
\en 
From   Eq.(12) we also find 
\be 
 \av{(F_{\rm bot}^x- F_{\rm top}^x)^2}_{\rm sim} 
= 4 C_{\rm pl}V/H^2 \quad (T\to 0). 
\en
On the other hand,    the kinetic stress    and the 
potential  stress deviation $\delta W_{\rm p}(t)$   decay rapidly in Eq.(61).  
Thus,  
\be 
C(0)- C_{\rm pl}= n (k_BT)^2+\langle  
\av{ {\delta W_{\rm p}}^2 }_\ih 
{\rangle}_{\rm sim}/V 
\en  
where  $n(k_BT)^2$ is   the kinetic contribution  
 with $n=N/V$  and $\av{ {\delta W_{\rm p}}^2 }_\ih $ is 
given in  Eq.(42) for each inherent state.

From Eqs.(40) and (64)     the  average of $\mu$ is written  as   
\be 
\av{\mu}_{\rm sim}
= (\av{\mu_\infty }_{\rm sim}+ nk_BT)-
(C(0) - C_{\rm pl})/k_BT,
\en 
which holds at finite $T$.  Since  Eq.(11) indicates 
$F_{\rm bot}^x=- F_{\rm top}^x$ at $T= 0$, 
  Eqs.(62)-(65) yield  
the  wall-force   correlation including the inherent 
contribution,   
\bea 
&&\hspace{-1.5cm}
\av{F_{\rm bot}^x F_{\rm top}^x}_{\rm sim} H^2/V =
k_BT\av{\mu}_{\rm sim}-C_{\rm pl} \nonumber\\
&& =  k_BT(\av{\mu_\infty }_{\rm sim}+ nk_BT) 
-C(0) .
\ena 

\begin{figure*}[t]
\includegraphics[scale=0.9]{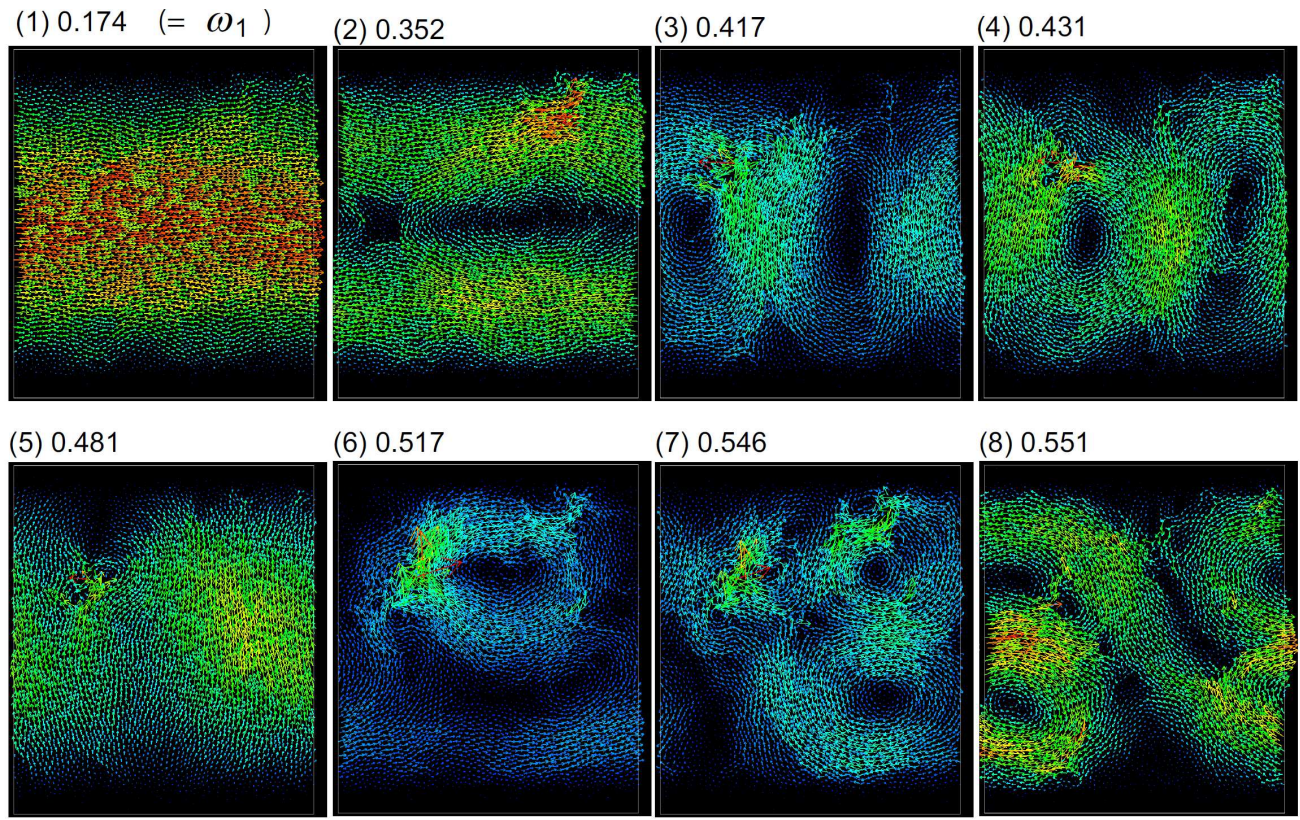}
\caption{ First eight    normal  modes $e_{i\alpha}^\lambda$ 
($1\le \lambda\le 8$) in two dimensions  
with  nearly rigid boundary walls at $z=\pm H/2$ and 
 the periodic boundary condition along the $x$ axis, which are  
obtained from  Eq.(54)  with 
the normalization $\sum_{i,\alpha} (e_{i\alpha}^\lambda)^2=1$. 
 Number on each panel  gives  $\omega_\lambda$. 
Colors of the particles  represent the displacement 
magnitude  $|{\bi u}_i|$.  The first and  second  modes 
correspond to the transverse sounds. 
The  fifth mode  is the first longitudinal sound. 
In the third, forth, and eighth  modes, 
transverse and longitudinal displacements are mixed 
due to the walls. In the third, sixth and seventh modes, 
quasi-localized  regions of large 
displacements can be seen in the upper left region. 
 }
\end{figure*}

\begin{figure*}
\includegraphics[width=0.96\linewidth]{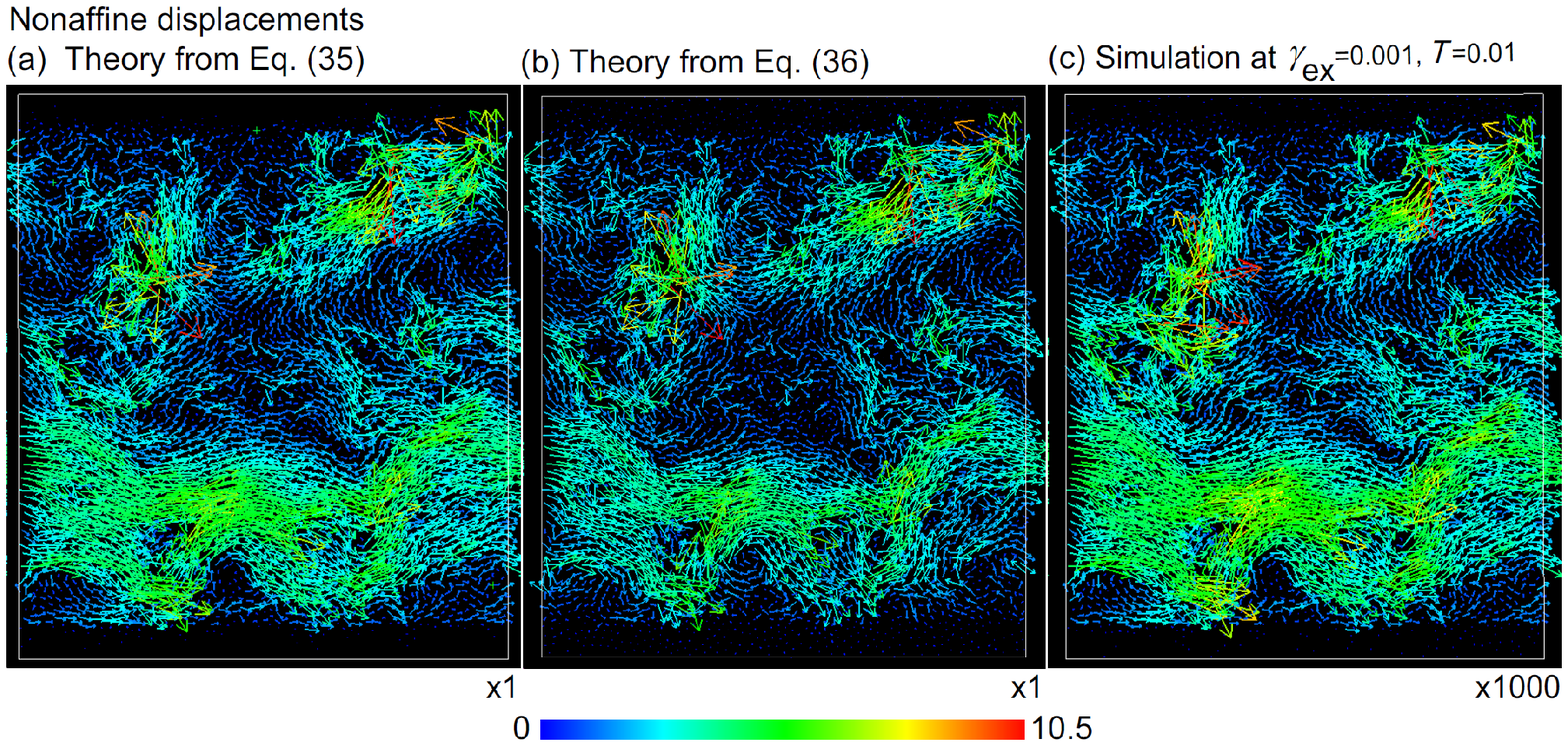}
\caption{(Color online)  Nonaffine displacements 
 $({\bar u}_i^x/\gamma_{\rm ex}-z_i, {\bar u}_i^z/\gamma_{\rm ex})$ 
 divided by  mean  strain $\gamma_{\rm ex}$ 
in the $x$-$z$ plane $(d=2)$ at low $T$, where those of 4000 unbound 
particles can be seen but those of 500 bound particles 
are  invisible. Colors represent  magnitudes of these vectors. 
They are calculated from  Eq.(35) in (a), from Eq.(36) in (b), 
and from simulation of applying a strain of $\gamma_{\rm ex}=10^{-3}$ 
at $T=0.01$ in (c). In (a) and (b),  
use is made of the  inverse Hessian matrix $(h^{-1})_{ij}^\ab$.  
In (c),  the particle displacements  
 are completely reversible  and 
the depicted positions are those  averaged  
 over a time interval of $5000$, where  thermal vibrational  motions 
are removed. Results in (a), (b), and (c) 
are very close to support  our theory.  
}
\end{figure*}

\section{Vanishing of shear modulus} 

\subsection{Supercooled  liquids}

In supercooled liquids,  the configuration changes occur  appreciably 
on   timescales longer than  the bond 
breakage time $\tau_b$\cite{Kawasaki,Yama1,Shiba}.    
In each plastic  event,   some bonds are broken and some particles  
 jump over distances longer than 
their  diameters.  As a result, 
 the diffusion  constants of the two components\cite{Kawasaki} 
become proportional to  $\tau_b^{-1}$ with 
the coefficients  independent of $T$.
Therefore,  the quasi-equilibrium distribution $P_\ih$ 
in Eq.(17) can  be used  on  timescales shorter   than $\tau_b$, 
where    the mean-square  displacements are smaller 
 than the square of the particle diameters.  
At higher   $T$, 
the second term in Eq.(65)   increases  
such that $\av{\mu}_{\rm sim}\to 0 $  
at a transition temperature\cite{Wil,Evans,Yoshi,Saw,Ilg,Teren}.

For   shear rate ${\dot\gamma}_{\rm ex}$, 
we have   a   Newtonian    regime 
for ${\dot\gamma}_{\rm ex}\tau_b<1$ and 
 a  shear-thinning regime  for 
  ${\dot\gamma}_{\rm ex}\tau_b>1$\cite{Kawasaki,Yama1,Shiba},
where  the  Newtonian viscosity  is of order $\mu\tau_b$ 
and  the nonlinear one is of order $\mu/{\dot\gamma}_{\rm ex}$. 
Note that   the Green-Kubo formula 
for the former depends    on the ensemble   
   for  deep supercooling\cite{Harro,Yip}, where 
the average    $\av{\cdots}_{\rm sim}$ in Eq.(61) 
has to be still   in  some limited phase-space region.

\subsection{Liquids}

In liquids at higher $T$,
thermal    equilibration  is rapidly   achieved within experimental times. 
We  can  use the linear response theory  around 
 the equilibrium  distribution  $P_{\rm eq}\propto \exp(-{\cal H}/k_BT)$ with 
 ${\cal H}= \sum_i |{\bi p}_i|^2/2m_i+ U$.  
Here,  the equilibrium average over $P_{\rm eq}$
 is written as  $\av{\cdots}_{\rm eq}$.

 If the  boundaries  are slightly moved, 
the  perturbed  Hamiltonian is 
${\cal H}'={\cal H}  -\gamma_{\rm ex}(t) {\cal A}$, 
where  $\cal A$ is given in  Eq.(9) with $\av{{\cal A}}_{\rm eq}=0$. 
For any variable ${\cal B}$,  its average response is thus  expressed in 
the form of Eq.(30) with  
\be 
\chi_{BA}(t)= \av{{\cal B}(t) {\cal A}(0)}_{\rm eq}/k_BT. 
\en  
 If  ${\cal B}(t) 
=J_x({\bi r},t)= \sum_i p_i^x(t) \delta ({\bi r}-{\bi r}_i(t))$ 
in Eq.(67),    the response  depends  on $z$ and $t$. 
The resultant  boundary flow profile will be investigated in future.

 In liquids, the static shear modulus $\mu$ vanishes  from 
   $\av{ W {\cal A}}_{\rm eq}=0$,  
 where   $W =\int d{\bi r}\Pi_{\rm xy}$ is the total 
shear stress. Using $\mu_\infty$ in Eq.(41) 
we   rewrite it  as  
\be 
\av{\mu_\infty}_{\rm eq}+nk_BT= \av{W^2}_{\rm eq}/Vk_BT,
\en 
where the  left hand side   is called 
the high-frequency shear modulus\cite{Zwanzig}.  
We  can  prove Eq.(68)  using the   relation\cite{Zwanzig},
\be 
\av{{\cal B} f_i^\alpha}_{\rm eq} 
= -k_BT \av{\p {\cal  B}/\p x_i^\alpha}_{\rm eq}.  
\en 

\begin{figure}
\includegraphics[width=0.96\linewidth]{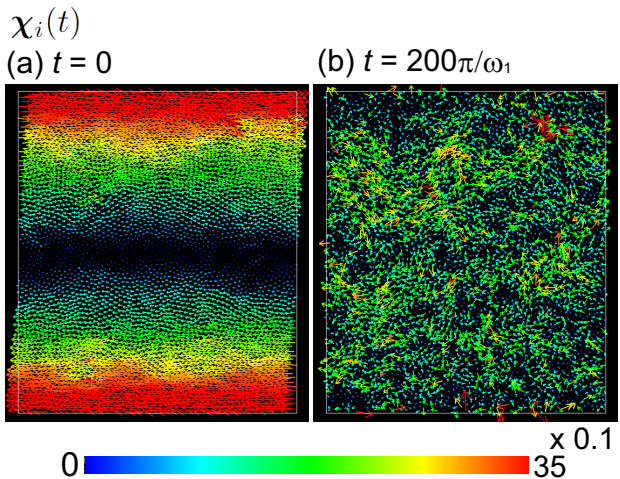}
\caption{(Color online) 
 Response functions 
$(\chi_i^x(t), \chi_i^z(t))$ in Eq.(76)  obtained from 
Eq.(35), 
 where (a) $t=0$ and (b)  $t=200\pi/\omega_1$. 
Colors represent magnitudes of these vectors. 
Those in (a) are equal to 
normalized displacements  $({\bar u}_i^x/\gamma_{\rm ex}, 
{\bar u}_i^z/\gamma_{\rm ex})$, where the affine parts 
are dominant  near the walls.   
In (b), small-amplitude vibrational modes can be seen, 
but the intial affine parts have disappeared.   
   }
\end{figure}

 For a small oscillatory 
shear rate ${\dot\gamma}_{\rm ex}= {\dot \gamma}_0\cos(\omega t)$, 
the average shear stress is  the real part of 
$-{\dot \gamma}_0 \eta^*(\omega) e^{{\rm i}\omega t}$.  
From  Eq.(10) we find the complex shear viscosity,  
\bea 
&&\hspace{-8mm}
\eta^*(\omega)= 
\frac{1}{Vk_BT}  \int_0^\infty dt e^{-{\rm i}\omega t}\Big[ 
\av{W(t) W(0)}_{\rm eq} \nonumber\\
&&\hspace{1cm} +{{\rm i}\omega} \av{W(t){\cal G}_x(0)}_{\rm eq}\Big],
\ena
where ${\cal G}_x (t)= \int d{\bi r} z J_x({\bi r},t)= 
\sum_j  z_j(t)  {p}_j^x(t)$.
At $\omega= 0$,   the Green-Kubo formula surely holds for the 
viscosity $\eta= 
\eta^*(0)$   in the bulk.   We can also 
use   the argument below Eq.(50) to fluids. 
For liquids,   Eq.(30) gives    
\be  
 \eta =\frac{1}{Vk_BT} H^2
\int_0^\infty dt  
\av{  F_{\rm top}^x (t)  F_{\rm bot}^x(0)}_{\rm eq}, 
\en    
which should be compared  with Eq.(50) for elastic films.  
 

\begin{figure*}[t]
\includegraphics[width=0.96\linewidth]{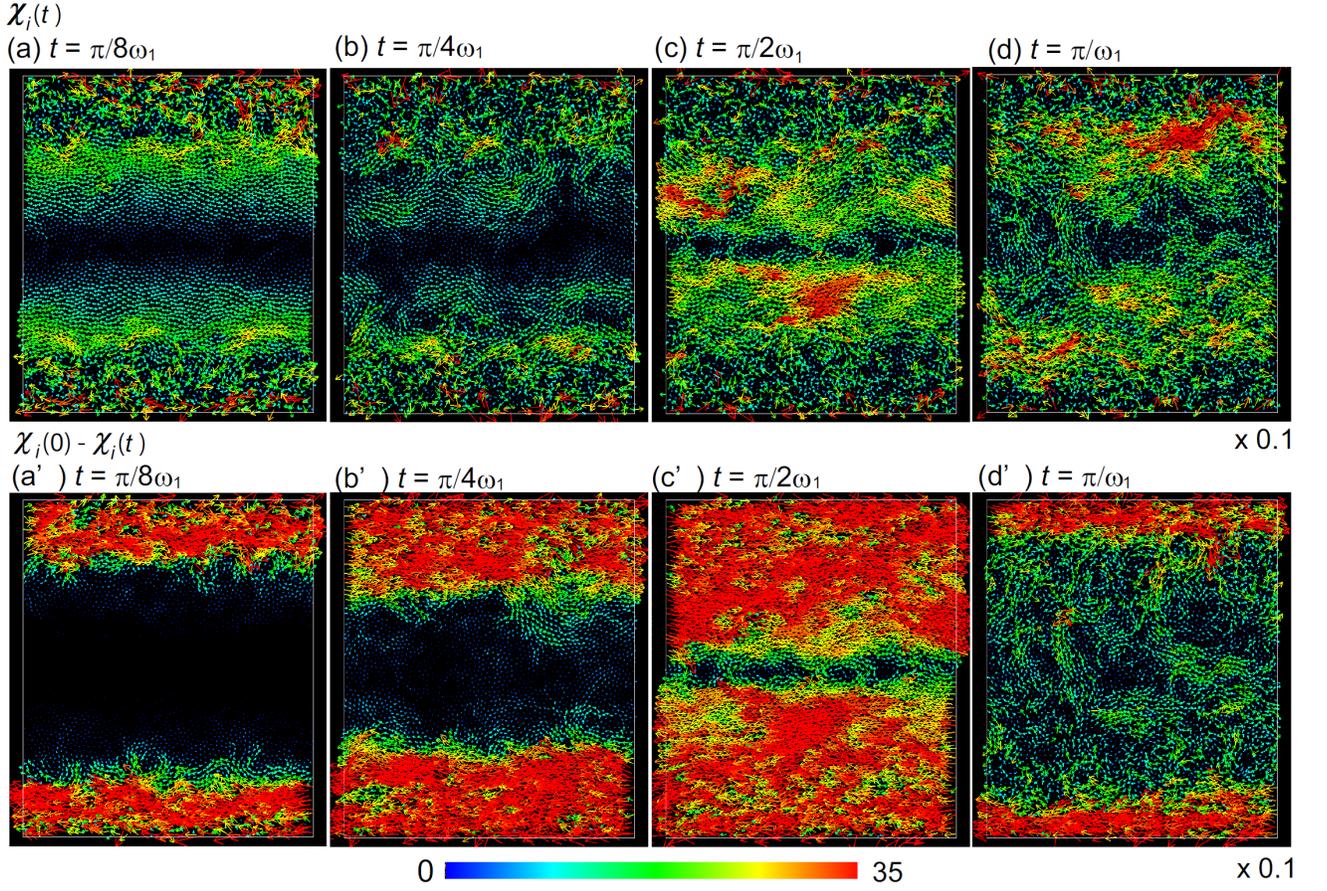}
\caption{(Color online) Time-dependent response functions  
${\boldsymbol \chi}_i(t)=   
(\chi_i^x(t), \chi_i^z(t))$ in Eq.(76) (top)  
and differences ${\boldsymbol \chi}_i(0)
- {\boldsymbol \chi}_i(t)$ in Eq.(77) 
(bottom)  in the $x$-$z$ plane $(d=2)$, where  
 $\omega_1 t$ is  $\pi/8$,  $\pi/4$,   $\pi/2$, and  $\pi$ 
(in the first  half period)  with $\omega_1=0.174 (\cong \pi c_\perp/H)$.  
Top: Affine correlation  at $t=0$ 
 disappears as  transverse sounds propagate inward with speed $c_\perp=3.9$ 
from the  walls. 
Bottom: These are normalized disturbances after  stepwise wall motions 
at $t=0$, which propagate from the walls and meet 
at $t=\pi/2\omega_1$.  Colors represent the magnitudes of these vectors. 
These are calculated from the linear-response relations in Sec.IIG.  }
\end{figure*}

\begin{figure*}
\includegraphics[width=0.9\linewidth]{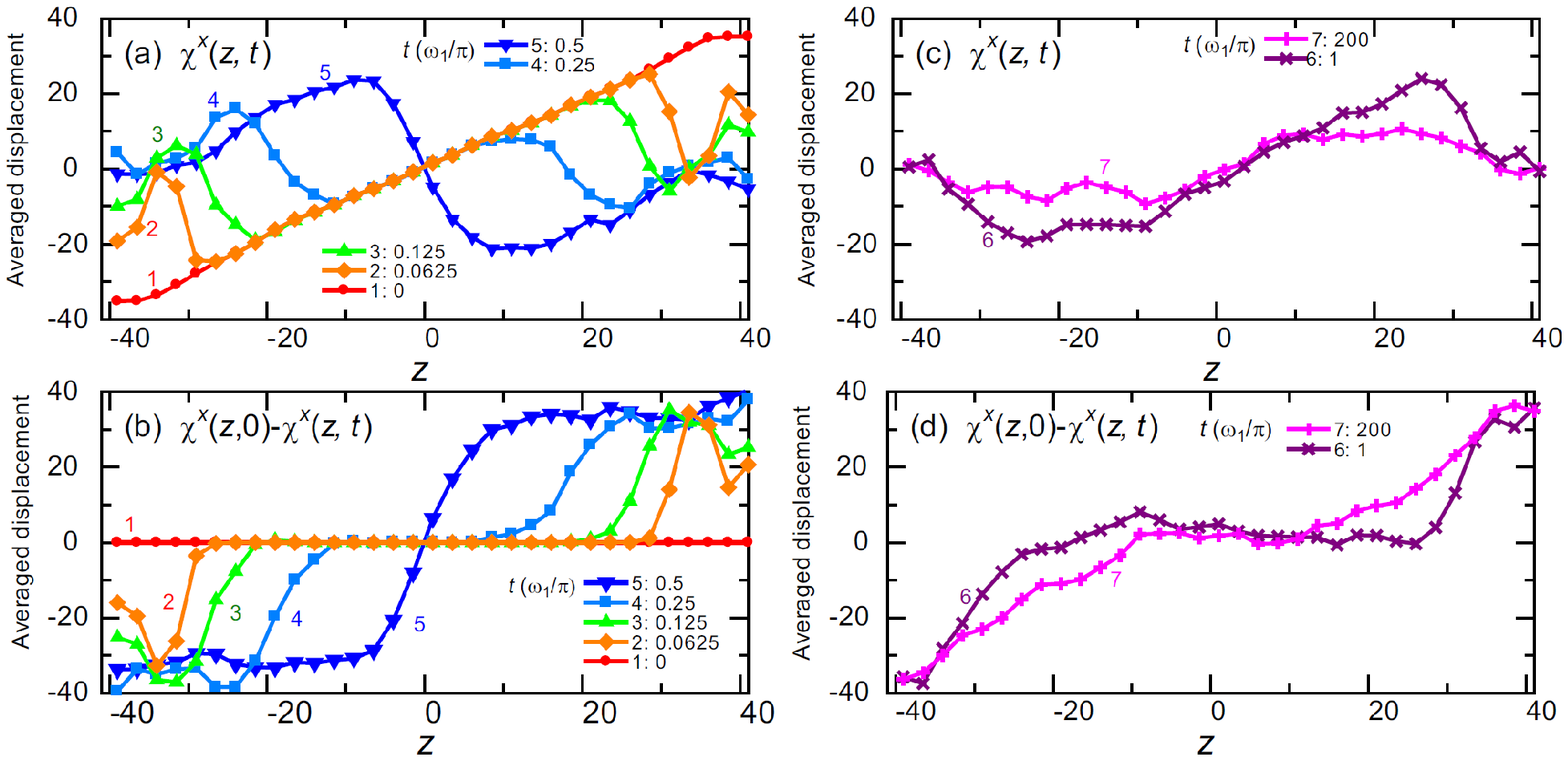}
\caption{(Color online) Laterally averaged 
response function (x component) ${\bar\chi}^x(z,t)$  (top) 
and ${\bar\chi}^x(z,0)-{\bar\chi}^x(z,t)$ (bottom) 
as functions of $z$. These are defined in Eq.(78)
 in the range $|z|<H/2+\ell_{\rm w}$. 
Normalized time 
 $\omega_1 t/\pi$ is  $0, 1/16, 1/8, 1/4,$ and $1/2$ in (a) and (b) 
and  is 1 and 200 in (c) and (d). In (a), the initial affine correlations  
disappear with propagation of transverse sounds from the walls. 
Relaxation at the walls takes place on a timescale of $\tau_{\rm w}\sim 2$. 
In (b), the  boundary displacements  propagate 
inward with $c_\perp$ as a shock wave with front thickness 
$c_\perp\tau_{\rm w}$.   In (c) and (d), the correlations 
decay very slowly at long times.  
}  
\end{figure*}

\section{Numerical results in two dimensions}

\subsection{Simulation method  } 

We now present numerical results in two dimensions 
($d=2$) in the $x$-$z$ plane.   
We have briefly  described  our model at the beginning of Sec.II. 
Our system   is composed of 
two particle species with numbers $N_1=N_2=N/2=2000$ in the cell.  
The pairwise   potentials are given by  
\be 
\phi_{ab} (r) = 
\epsilon  {[(\sigma_{a}+\sigma_b)/2]}^{12} [{r}^{-12} -r_c^{-12}]  
\quad (r<r_{\rm c}), 
\en 
where $a$ and $b$ represent the particle species 
and we  introduce   $\epsilon$, $\sigma_{1}$, and  $\sigma_2=1.4\sigma_1$. 
The potentials $\phi_{ab} (r) $  
 vanish for $r\ge r_{\rm c} =2.25( \sigma_a+ \sigma_b)$.  
The   mass ratio is  $m_2/m_1=1.96$. The  cell lengths  are  
 $H=L=70.2\sigma_1$. The average particle density 
is $n=N/LH= 0.81 \sigma_1^{-2}$ and the average mass density 
is $\rho= 1.2m_1\sigma_1^{-2}$.  
Hereafter, we will measure 
space, time, and temperature   in units of $\sigma_1$, 
$t_0= \sigma_1(m_1/\epsilon)^{1/2}$, and $\epsilon/k_B$, respectively.

As in Fig.1, we attach  two boundary layers 
with thickness  $\ell_{\rm w}= H/16$. Each layer contains 
$M=250$  particles  bound  to  pinning points   ${\bi R}_j$  
 by   the  potential (2). The total particle number is  
$N+2M=4500$.  The    ${\bi R}_j$ were  
particle positions  in a liquid  state\cite{Shiba}.
The  spring constant is chosen to be  large at $s_0 = 200$. 
Then, the displacements of the bound particles from their inherent positions 
$u_k^\alpha$ ($k>N$) 
 undergo thermal motions with   amplitudes  of order $(T/s_0)^{1/2}=0.07
T^{1/2}\ll 1$.
Under applied strain, the motions of the bound particles 
are very small $(\propto s_0^{-1}$), which 
become even smaller with increasing the distance from the cell region. 
Our walls are  nearly 
 rigid   and their surfaces are  rough  as in  Fig.1, so  
slip motions do not occur  for small wall motions. 


We followed the following steps. 
(i) We started with  a liquid at  high $T$, 
  lowered $T$  to 0.01 without crystallization,  
and waited  for a time of  $10^3$ to  
realize a   glassy  state,
 where  we  attached   Nos\'e-Hoover thermostats in the cell 
and in the boundary layers\cite{Shiba}.  
(ii) We carried out some simulation runs 
at $T=0.01$  removing  the thermostats. 
Results from  these runs will be 
presented  in Fig.3(c) and Fig.7(a). 	
(iii) To seek  an  inherent state, 
we further   cooled our  system  down to $T=10^{-5}$ 
keeping the thermostats without applied strain. 
Then,  the particle positions became frozen.  
We use the resultant  inherent state  in Figs.2-9   
without taking  the ensemble average. 

\subsection{Inherent state and eigenmodes}

First, we   examine the validity of 
 Eqs.(62) and (63) in the limit  $T\to 0$.    Indeed, we obtain  
$(W^\ih)^2/V=31.1$ and $(H^2/4V){(F_{\rm bot}^x- F_{\rm top}^x)^2}=
32.2$, while we find the 
plateau value  $C_{\rm pl}=27.7$   from 
a simulation run at $T=0.01$.  
These three values are fairly  close even for a single 
inherent state.  For the inherent particle positions, 
we calculated  the coefficients 
$w_i^\alpha$ in Eq.(26),  the   Hessian matrix 
$h_{ij}^{\alpha\beta}$ in Eq.(19), and its  inverse $(h^{-1})_{ij}^\ab$ 
satisfying Eq.(20). The   product of these two matrices 
is very close to the unit matrix 
$\delta_\ab \delta_{ij}$   with 
differences of  order $10^{-16}$ for all the elements.  
The shear modulus $\mu$ is then given by 
$18.1$ from Eq.(39), 
 $15.4$ from Eq.(40), 
$17.4$ from Eqs.(44) and (50), while it is 
$16.0$ from the stress-strain relation obtained from a 
simulation in Fig.3(c). 


In Fig.2, we display  the 
first eight  eigenvectors $e_{i\alpha}^\lambda$,  
where   the displacements of  the bound particles are 
very small and are invisible here.    The 
 lowest eigenfrequency  is $\omega_1=0.174$. This corresponds to  
  the  first   transverse sound mode $e_{i\alpha}^1$, 
which is roughly proportional to  
$ \delta_{\alpha x} 
\sqrt{m_i} \cos(\pi z_i/H)$ $(i\le N)$.  
If we set  $\omega_1= \pi c_\perp /H$, 
we obtain the transverse sound speed   $c_\perp =  3.9$. 
The same  speed   follows 
from  $c_\perp = (\mu/\rho)^{1/2}$ for 
  $\mu=16$ and  $\rho=1.2$.  
The  first longitudinal mode appears at $\lambda=5$ 
with $\omega_5= 0.481$.  
We also obtain strongly  localized modes 
for $\lambda=6$ and $ 7$, where  
clusters of large-amplitude oscillation 
are weakly connected to the bulk\cite{Sch}.

We  project  
 the variables $\delta W_{\rm p}$
in Eq.(25) and $\delta{\cal A}$ in Eq.(27) on the eigenmodes as 
$\delta W_{\rm p}=\sum_\lambda Z_W^\lambda s_\lambda$ 
and $\delta{\cal A}=\sum_\lambda Z_A^\lambda s_\lambda$ (see Sec.IIG). Here, Eq.(58) gives  
\be 
 Z_W^\lambda= -\sum_{i,\alpha} \frac{w_i^\alpha e_{i\alpha}^\lambda }
{\sqrt{m_i}}, 
~Z_A^\lambda =   (\sum_{k\in {\rm bot} } -
\sum_{ k\in {\rm top} }) \frac{Hs_0e_{kx}^\lambda}{2\sqrt{m_k}}. 
\en 
From   Eq.(38),  
 $\mu$ is expressed as  
\be 
\mu= -\frac{
\av{\delta W_{\rm p} \delta {\cal A}}_\ih}{k_BT V}= - 
\sum_\lambda \frac{Z_W^\lambda Z_A^\lambda}{\omega_\lambda^2 V},  
\en  
which is calculated  to be   17.7.  
In Table 1, we give $\omega_\lambda$, $Z_W^\lambda$, and $Z_A^\lambda$ 
($\lambda\le 8$) for the inherent state 
under consideration, where  $Z_A^\lambda$  have 
 large amplitudes as compared to    $Z_W^\lambda$. In fact, 
we find   $\av{(\delta {\cal A})^2}_\ih /k_BT = 
\sum_\lambda (Z_A^\lambda)^2/\omega_\lambda^2= 2.5\times 10^4V$ from 
  Eq.(43), while we calculate  $\av{(\delta {W}_{\rm p})^2}_\ih /k_BT = 
\sum_\lambda (Z_W^\lambda)^2/\omega_\lambda^2= 21.2V$.
  Notice that Eq.(26) 
also gives $
 Z_W^\lambda= -\sum_{i,j, \alpha} w_{ij}^{zx\alpha}
E_{ij\alpha}^\lambda/2$, 
where $E_{ij\alpha}^\Lambda\equiv 
 e_{i\alpha}^\lambda/{\sqrt{m_i}}-  e_{j\alpha}^\lambda/{\sqrt{m_j}}$ 
are  small for most adjacent $i$ and $j$.   
See Eq.(B8) for the eigenmode 
projection of $\delta{\cal A}$ in the continuum elasticity.

\begin{table}
\caption{Vibrational eigenmodes and 
projection coefficients. }
\label{tab:3d}
\begin{tabular}{cccccccccc}
\hline 
${\lambda}$  && 1  & 2 & 3 & 4 & 5 & 6 & 7 & 8 \\
\hline
$\omega_{\lambda}$  && 0.174 & 0.352 & 0.417 & 0.431 & 0.481 & 0.517 & 0.546 & 0.551\\
\hline
$Z_W^\lambda$  && 1.375 & 11.76 &-0.226 & -1.730 & -7.920  & -9.957 & 11.71 
&-13.64 \\
\hline
$Z_A^\lambda$  && 37.83  & -66.11 & 24.41 & -13.08 & 231.3 & -153.7 & 118.7 
& 26.22 \\
\hline
\end{tabular}
\end{table}

\subsection{Nonaffine displacements at static strain}
 
In Fig.3, we display the normalized  nonaffine displacements 
 $({\bar u}_i^x/\gamma_{\rm ex}- z_i, {\bar u}_i^z/\gamma_{\rm ex})$ 
for all the particles. 
They   are  highly heterogeneous on  large scales 
for  the unbound particles, as has been reported in the 
literature\cite{PabloP,Malo,Malo2,Barr1,Barr2,Miz1,Liu,Wit,Wit1}. 
Those  of the bound particles are  small and invisible.
To confirm the validity of our theory, 
we  calculated these results (a) from the wall-particle correlations 
in  Eq.(35), (b) from the particle-particle 
correlations in  Eq.(36), 
and (c) from a single run of  molecular dynamics  simulation of 
 applying a strain of $\gamma_{\rm ex}=10^{-3}$  at $T=0.01$.  
In (a) and (b), 
we  use  the  inverse Hessian matrix 
and the results   are the  averages  over $P_\ih$ in Eq.(17). 
In (c), use is made of  the common  inherent state,  
  the time average is over  a time interval of $t_{\rm sim}= 5000$, 
and there is no irreversible motion.   We can see  
  good agreement of the  results in (a), (b), and (c)  from 
the three methods, which supports our theory. In particular,   those  
in  (a) and (c) are very close.

In  Fig.4(a), we present   the full normalized displacements 
 $({\bar u}_i^x/\gamma_{\rm ex}, {\bar u}_i^z/\gamma_{\rm ex})$ 
using  Eq.(35), whose  affine parts are   conspicuous 
 near the walls. Here,  Eq.(60) gives 
\be 
Z_A^\lambda= \omega_\lambda^2\sum_{i,\alpha}
(e_{i\alpha}^\lambda/{\sqrt{m_i}})
({{\bar u}_i^\alpha}/{\gamma_{\rm ex}})  ,
\en 
which is equivalent to  the second relation in  Eq.(73).    
If we set   ${{\bar u}_i^\alpha}/{\gamma_{\rm ex}} 
=z_i \delta_{\alpha x}$ in Eq.(75), 
we obtain the affine part of  $Z_A^\lambda$. 
For  $\lambda=1$, it is  
$-0.54$, so $Z_A^1$ mostly 
consisits of   the nonaffine part 
since $Z_A^1\cong 38$ (see Eq.(B8)).

\subsection{Space-dependent  dynamics}


  Next, we examine space-time-dependent effects. For 
    $\delta{\cal B}= u_i^\alpha$ in Eq.(59), 
 we define the  response functions,   
\be
\chi_i^\alpha(t) =  
 \frac{\av{u_i^\alpha (t)\delta {\cal A}(0)}_\ih}{k_BT}
=  \sum_\lambda  
\frac{e_{i\alpha}^\lambda }{\sqrt{m_i}}
Z_A^\lambda
\frac{\cos(\omega_\lambda  t)}{\omega_\lambda^2}, 
\en 
for all the particles. 
At $t=0$, we have $\chi^\alpha_i(0)= {\bar u}_i^\alpha/\gamma_{\rm ex}$ 
for  a static strain   in Fig.4(a).  
We also show  $\chi_i^\alpha(t) $ at $t=200\pi/\omega_1$ 
in Fig.4(b), which 
   retains  no affine part but still keeps   some  
space correlations.  
For   a stepwise strain $\gamma_{\rm ex}(t)=\gamma_0\theta(t)$, 
which is  zero for $t<0$ and is a constant $\gamma_0  $ 
for $t>0$,   Eq.(33) yields the subsequent  evolution,   
\be 
{\bar u}_i^\alpha(t)/\gamma_0=
\chi_i^\alpha(0)-\chi_i^\alpha(t)\quad (t>0).
\en

In Fig.5,  we show $\chi_i^\alpha(t) $ in the upper panels 
and $\chi_i^\alpha(0)-\chi_i^\alpha(t) $ 
in the lower panels for $\omega_1 t/\pi =1/8, 1/4, 1/2$, and 1,  In the initial stage, 
disturbances advance  from the walls with the  speed $c_\perp$ 
without noticeable  changes in the center region. 
 In (a)-(c),  the initial affine correlations 
 in Fig.4(a) disappear from the walls. 
In  (a')-(c'), shock-like transverse sounds  propagate 
from the walls. Their fronts are  irregular due to random scattering. 
 In (d) and (d'), 
the sounds from the walls encounter at the center.

In Fig.6, we  display    the laterally averaged profiles,  
\be 
{\bar\chi}^x(z,t)= \frac{H}{N\Delta z}
\sum_i \theta({\Delta z/2-|z_i-z|}) \chi_i^x(t), 
\en   
where     $\theta(p)$ is the step function being 1 for $p>0$ 
and 0 for $p\le 1$.
We here remove the  glassy irregularities to 
 examine the acoustic  behavior and the boundary relaxation 
along the $z$ axis.  Setting  $\Delta z=2.5$, 
we plot (a) ${\bar\chi}^x(z,t)$   and (b) ${\bar\chi}^x(z,0)- 
{\bar\chi}^x(z,t)$, where 
$\omega_1 t/\pi= t/18=0, 1/16, 1/8, 1/4,$ and $1/2$. 
In (a), the initial  affine 
correlation soon   disappears  near the walls, 
whose  timescale  $t_{\rm w}$ is about  $2$. 
 In (b),  the boundary values  of 
${\bar\chi}^x(z,0)- {\bar\chi}^x(z,t)$ at $z\cong \pm H/2$ 
change  from 0 to the static values ($\sim \pm 40$) on 
the  time $t_{\rm w}$.  
In the initial stage $t\ll t_{\rm a}= H/c_\perp\cong  18$, 
the expanding  sound from each wall 
 is  of the form $g( t-\ell/c_\perp)$, where $\ell$ is 
the distance  from  the wall and $g(t)$ is the boundary value being 
zero for $t\le 0$.  Since 
$t_{\rm w}\ll t_{\rm a}$, 
  a  shock wave is produced  from each wall, whose  front  
has a thickness  of order $c_\perp t_{\rm w}\sim 10$.  
 In (c) and (d), we show the profiles at 
 $\omega_1 t/\pi =1$ and 200, where complex  oscillations 
still remain nonvanishing. 

In agreement with 
  our results,  Jia {\it et al.}\cite{Jia}   observed  a coherent 
ballistic pulse  and  speckle-like   signals  
in   sound propagation  in a granular matter. 
Their  observation was later reproduced 
 in a simulation of  granular matter\cite{Roux}.


\subsection{Time-correlation functions}

\begin{figure}
\includegraphics[width=0.96\linewidth]{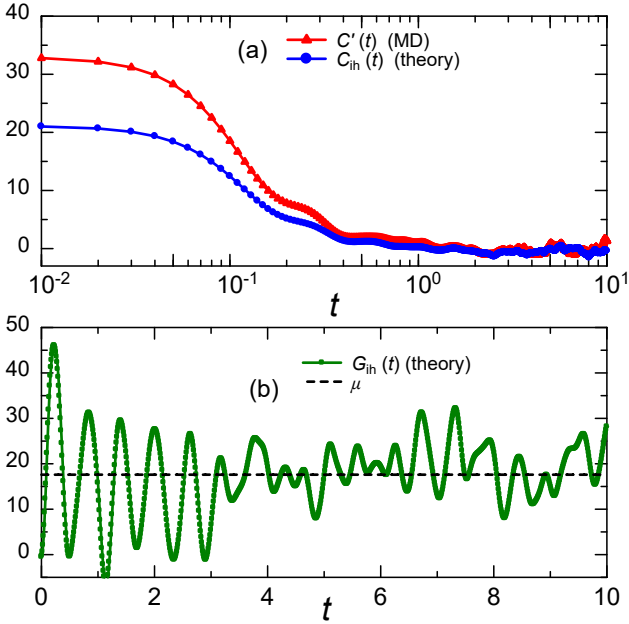}
\caption{(Color online) 
(a)  $C_\ih (t) $ in Eq.(79) and $C'(t)\equiv  [C(t)-C_{\rm pl}]/k_BT$ 
in Eq.(80) vs $t$ on a semi-logarithmic scale, where the former is 
from time-evolution of the eigenmodes  and the latter is from molecular dynamics simulation at $T=0.01$. These decay on a timescale of 0.2. 
(b)  $G_\ih (t) $ in Eqs.(34) and (81) vs $t$, 
which is zero at $t=0$, has  a maximum ($\sim 47)$ 
at $t=0.23$, and exhibits 
complex oscillations around $\mu\cong 18$.   
   }
\end{figure}

\begin{figure}
\includegraphics[width=0.9\linewidth]{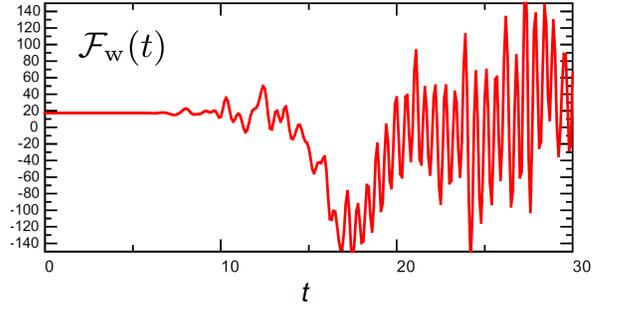}
\caption{(Color online)
Time-correlation function ${\cal F}_{\rm w}(t)$ 
in Eq.(82) for the forces from the walls to the particles 
as a function of $t$. 
At short times, it  equal to $17$. 
For  $t\gs t_{\rm a}= H/c_\perp\cong 18$, it fluctuates 
irregularly due to scattering.  
}  
\end{figure}

In Fig.7(a), we examine  the stress time-correlation function  
at a fixed inherent state. We write it as       
\be
C_\ih (t) = \frac{\av{\delta W_{\rm p}(t) \delta{ W}_{\rm p}(0)}_\ih}{k_BT V} 
=\sum_\lambda  \frac{
(Z_W^\lambda)^2  }{V\omega_\lambda^2} 
{ \cos(\omega_\lambda t)} . 
\en 
In (a), this function  decays to nearly zero on a rapid  timescale of 0.2. 
Its   oscillatory behavior is suppressed because of relatively 
small $Z_W^\lambda$  for not large  $\lambda$ in Table 1, where 
the first term in Eq.(79) is $0.013 \cos(\omega_1 t)$. 
Notice that the large-scale  sound modes 
do not contribute significantly  to $\delta W_{\rm p}(t)$,  
because  the difference $u_{ij}^\alpha=u_{i}^\alpha- u_{j}^\alpha$ 
appears for adjacent $i$ and $j$ in the first line of  Eq.(25).

We also calculated  the stress time-correlation function  
$C(t)$ in Eq.(61) at $T=0.01$ in a single run 
of molecular dynamic simulation with the same inherent state. Here,   
 the time average was 
taken over the simulation time  $t_{\rm sim}=5\times 10^4$. 
We used  the stress integral $W(t)$ in Eq.(61) including  the 
contributions from the bound particles, 
so there should be some  boundary effect. 
Since  Eq.(64) is predicted,  Fig.7(a) gives  
\be 
   C'(t)=  [C(t)-C_{\rm pl}]/k_BT,
\en  
where  $ C_{\rm pl}=27.7$ is the plateau value. 
In (a),  its  initial value  
$C' (0) =33.1$ is  larger than $C_\ih (0) =21.2$, 
but  the two curves   fairly agree for $t\gs 0.1$.

In Fig.7(b), we plot  the stress relaxation function $G_\ih(t)$  
in Eq.(34),  which can be expressed as 
\be 
G_\ih(t)=- \sum_\lambda  \frac{
Z_W^\lambda Z_A^\lambda  }{V\omega_\lambda^2} \Big[
{1- \cos(\omega_\lambda t)} \Big].
\en 
In (b), this function   exhibits complex oscillatory behavior 
with timescales  shorter than $t_{\rm a}= c_\perp/H\sim 18$.

In Fig.8, we   examine   the time-correlation function 
of the forces from the walls. From Eq.(50), it is scaled as 
\be   
{\cal F}_{\rm w}(t)= 
\av{\delta F_{\rm top}(t)\delta F_{\rm bot}(0)}_\ih H^2/Vk_BT.
\en  
This function can be expressed in the form of Eq.(59) 
from Eqs.(28) and (58). 
Starting with   ${\cal F}_{\rm w}(0)=\mu \cong 17$. 
  ${\cal F}_{\rm w}(t)$  is nearly constant for 
 $t< t_{\rm a}\sim  18$. 
Around $t\sim t_{\rm a}$, the sound from the bottom  
is reflected in the reverse direction at the top, 
resulting in a large drop in  $\delta F_{\rm top}(t)$. 
For $t \gs t_a$, it largely fluctuates due to scattered waves. 
See Appendix B for   ${\cal F}_{\rm w}(t)$ for 
homogeneous $\mu$.

From the argument below Eq.(50) we recognize  
that  ${\cal F}_{\rm w}(t)$ in Eq.(82)  is  measurable  experimentally. 
That is, at $t=0$, we  move the bottom layer by $-H \gamma_0$ in a 
stepwise manner keeping  the top layer  at rest. Then, from  Eq.(33), 
  the average  force from  the top wall 
to the particles   is given by 
$\gamma_0[{\cal F}_{\rm w}(0)-{\cal F}_{\rm w}(t)]$ 
per unit area  at time $t(>0)$. 
As in Fig.8, it should  vanish before 
arrival of the sounds.  

Somfai {\it et al.}\cite{Roux} 
 calculated  a force signal on a wall 
after emission of  a pulse strain  from the 
opposite wall in a granular model. Their signals  
without damping resemble  ours  in Fig.8, but they decay to zero with 
increasing  viscous dissipation. 
Wittmer {\it et al.}\cite{Wit1}    
introduced   {  friction}  in  
 the stress relaxation in a random elastic  
network.

\begin{figure}
\includegraphics[width=1\linewidth]{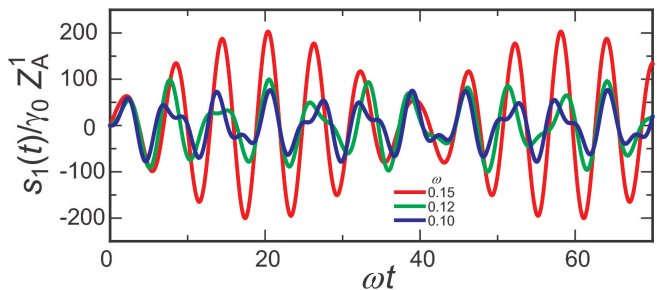}
\caption{(Color online) Average 
time-evolution of the first-mode amplitude ${\bar s}_1(t)$ 
in Eq.(83) (divided by $\gamma_0 Z_A^1$)  after application of 
sinusoidal  strain $\gamma_{\rm ex}(t)=\gamma_0\cos(\omega t)$ for $t>0$, 
where  $\omega  /\omega_1=0.1, 0.12$, and 0.15.
}
\end{figure}

\subsection{Resonance in periodic straining} 

We also suppose  application of a sinusoidal strain 
$\gamma_{\rm ex}(t)=\gamma_0 \theta (t) \sin(\omega t)$, 
which is zero for $t<0$.   For $t>0$,  
substitution of this strain into  Eq.(30) yields  
the average displacements 
${\bar u}_i^\alpha(t) = \sum_\lambda e_{i\alpha}^\lambda 
{{\bar {s}_\lambda}(t)}/\sqrt{m_i}$. Here, ${{\bar {s}_\lambda}(t)}$ 
is the time-dependent average of the mode amplitude
 $s_\lambda(t)$ in Eq.(55). Then, ${{\bar {s}_\lambda}(t)}$ 
is a mixture of two oscillations, 
\be 
{{\bar {s}_\lambda}(t)}
= \frac{{\gamma_0} Z_A^\lambda }{\omega_\lambda^2-\omega^2} \Big
[\sin (\omega t)-\frac{\omega}{\omega_\lambda}\sin (\omega_\lambda t)\Big] 
\quad (t>0). 
\en  
The  term with the intrinsic frequency $\omega_\lambda $ 
remains nonvanishing, but it decays to zero  in the presence of dissipation. 
Here,   ${\bar {s}_1}(t)$ for the first mode 
grows as $(\omega_1^2-\omega^2)^{-1}$ 
with increasing  $\omega$ towards  $\omega_1$.  
Thus, in  Fig.9,  ${\bar {s}_1}(t)$  is shown to 
increase with increasing  $\omega$. However, 
as $\omega$   approaches $\omega_1$, the local strain increases near the walls 
  and  the linear theory becomes invalid,

The above   displacement growth  
  is a resonance effect\cite{Landau1}.   Recently,  we have performed 
 a molecular dynamics simulation of a glass  
under periodic shear\cite{periodic}, where 
 plastic events are  proliferated at  resonance frequencies. 
 Wittmer {\it et al}\cite{Wit1} studied 
resonance in a model network, where the  
growth is suppressed  by the viscous friction.

\section{Summary and Remarks}

In Sec.II, we have presented a linear response theory 
of applying a mean  shear strain from boundary walls as in real 
experiments.  Our theory  is based on the relations (9) and (10) 
and  is  applicable to any solids  and fluids 
in confined geometries.  It can describe 
the linear dynamics  in the  
bulk and near the boundary walls in terms of the appropriate 
time-correlation functions. It  can be used 
even when the cell interior is   inhomogeneous. 
As a first nontrivial application, 
this paper has mostly  treated  
the linear response in glasses around a fixed  inherent state. 

In Sec.III, we have discussed 
the linear response in supercooled liquids 
with slow dynamics and in ordinary liquids 
with fast relaxations. In these states, 
 the viscosity comes into play in the bulk.  
Our theory  can further be used  
to study the boundary flow effects 
microscopically.

In  Sec.IV, we have presented  numerical results 
in a two-dimensional model  glass  on the basis of our theory. 
In Figs.3 and 4,  the forces from the walls 
are correlated with  the   displacements of all the particles 
in the cell, resulting in  heterogeneous responses. 
In Figs.5-9, time-dependent responses and time-correlation functions 
are strongly influenced by sound wave propagation 
and are very singular in glasses in the film geometry.

We make some remarks as follows.\\ 
(i)   In real  film systems, their  temperature 
is regulated by  heat transport  between 
the  walls  and the interior particles. 
To realize this situation in molecular dynamics simulation, 
we can  attach  heat bathes to  the boundary layers 
(not to the interior particles)\cite{Shiba}. 
Sounds are then  damped  upon  reflection at the walls, 
which leads to  decays of the time-correlation functions.\\  
(ii)  Plastic events   emit  sounds 
leading to fast transport of the released potential  energies  
throughout the system.  Shiba and one of 
the present authors\cite{Shiba} 
found that  plastic events cause large  oscillatory deviations  
in the force difference $ F_{\rm top}^x(t)- F_{\rm bot}^x(t)$, 
where the heat bathes in the boundary layers damp such oscillations.\\ 
(iii)  We can  construct a   microscopic theory of 
applying  dilational strains  by moving the 
walls along the $z$ axis, 
where propagation of the longitudinal sounds 
is crucial.  In particular,  the correlation between 
 the normal components $\delta F_{\rm top}^z$ 
and $\delta F_{\rm bot}^z$ can be 
obtained if  $\mu$  is replaced 
by $B+ (2-2/d)\mu$ in Eq.(50), 
where $B$ is the bulk modulus.\\
(iv) In future, we should  use a 
more realistic model of solid   walls,   
where  the forces from  the particles in the walls 
to those in the cell are of great importance\cite{slip}. 
 As discussed in Sec.IIIB, we can apply  our theory to fluids 
to investigate the boundary flow profiles.\\
(v)  There are a number of   elastic  systems with   inhomogeneous  elastic 
moduli on mesoscopic scales\cite{Head,Onukibook,Hecke,Luben,Wit1,Bas}, 
where  nonaffine strains  appear in  applied stress. 
In gels\cite{Head,Onukibook}, the crosslink structure 
is intrinsically  random depending on the preparation condition.
In  multi-component  metallic alloys\cite{Onukibook}, the elasticity  
in the presence of  precipitates is 
of great technological importance. which are harder  or softer than the matrix.


\acknowledgments{
This work was partially supported by the Japan Society 
for the Promotion of Science Grants-in-Aid for Scientific
Research (KAKENHI) 
(grants No. 16H04025, No. 16H04034, No. 16H06018, and 18H01188).}

\vspace{2mm}
\noindent{\bf Appendix A: One-dimensional 
 elastic systems}\\
\setcounter{equation}{0}
\renewcommand{\theequation}{A\arabic{equation}}
  
Here, we examine random  elastic  systems 
in one dimension in our theoretical scheme. 
We apply  a dilational strain $\epsilon_{\rm ex}$ 
  from  the walls at  low  $T$ to obtain an analytic 
expression for the dilational elastic  modulus. 
We can mention a number of random network models\cite{Wit1,Hecke,Luben}.

 Along the $x$ axis, 
the particles are at $x_i$ ($1\le i \le N) $ and 
the walls  are at $R_1$ and $R_{N}$, where  
$R_1<x_1<\cdots <x_N<R_N$ and $N\gg 1$.  We set $R_N=H/2$ 
and $R_1= -H/2$ with $H$ being the  system length. 
The end particles 1 and N are  bound 
to the walls by  potentials $\psi(d_0)$ 
and $\psi(d_N)$,  where 
 $d_0=x_1-R_1$ and    $d_N=R_N-x_N $.  
Particles $i$ and $i+1$ interact 
with potentials $\phi_i(d_i)$, 
where  $d_i=x_{i+1}- x_i$.
The total potential energy  is given by 
\be 
U=  {\sum_{i}}' \phi_i (d_{i})+\psi(d_0)+ \psi(d_N). 
\en  
The potentials  $\phi_i(d_i)$ can be random depending on $i$. 
Hereafter,  we set ${\sum_{i}}'=  \sum_{1\le i<N}$. 
The potential part of the microscopic pressure  is of the form,  
\be
 \Pi^{\rm p}(x) =-{\sum_{i}}' \phi_i'(d_i)\theta_i(x), 
\en 
where $\phi_i'(x)= d\phi_i(x)/dx$ and     
\be 
\theta_i (x) = 
{{\theta}}(x- x_{i}) -\theta(x- x_{i+1}).
\en 
Here,  $\theta(x)$ is  the step function  equal to 1 for $x>0$ 
and to 0 for $x\le 0$. Thus,  $\theta_i (x) $ is 1  in 
the interval   $x_i<x<x_{i+1}$ and is zero outside it.
From  Eq.(7) we find  
 ${\hat{\delta}}(x, x_i, x_{i+1})
=\theta_i (x)/d_i$, leading to Eq.(A2).

At fixed $H=R_N-R_1$, we assume that the 
  mechanical equilibrium holds at   $x_i= x_i^{\ih}$ as 
\be 
 \phi'_i (d_i^\ih)=\psi'(d_0^\ih)= \psi'(d_N^\ih) \equiv -p_\ih. 
\en  
where   $\psi'(x)= d\psi(x)/dx$ and  
$d_i^\ih$, $d_0^{\rm ih}$,  and $d_N^\ih(=d_0^\ih)$ are 
the equilibrium values  of $d_i$, $d_0$,  and $d_N$, 
respectively.   This state corresponds 
to   the { inherent  state} in glasses. 
 From Eq.(A4) the potential part of the   
pressure   is $\Pi^{\rm p}= p_\ih$ for  $x_{1}^\ih<x< x_{N}^\ih$. 
We next  consider   small displacements $u_i= x_i- x_i^\ih$. At fixed $H$, 
the deviation of $U$   is bilinear in $u_i$ to leading order  as 
\be 
\delta U= \frac{1}{2} {\sum_{i}}' 
{s_i} \xi_i^2 + \frac{1}{2} {s_{0}} (u_1^2+  u_N^2), 
\en  
where $\xi_i=d_i-d_i^{\rm ih}= 
 u_{i+1}- u_i$,  ${\sum_i}'  \xi_i= u_N-u_1$, and 
\be  
s_i=\phi_i''(d_i^\ih) \quad (1\le i<N), \quad 
s_{0} =\psi''(d_0^\ih),
\en   
with $\phi_i''(x) = d^2\phi_i (x) /dx^2 $ 
and $\psi''(x) = d^2\psi (x) /dx^2 $. We assume $s_i>0$ 
and $s_0>0$.  
At low   $T$,   $u_i$  fluctuate  thermally  obeying 
 the Gaussian distribution 
$P_{\rm ih}  \propto \exp(-\delta U/k_BT)$. 
Hereafter,  $\av{\cdots}_\ih$ denotes 
 this  average.

To linear order,  the deviation of $\Pi^{\rm p}(x)$ 
is written as 
\bea \hspace{-5mm}
\delta \Pi^{\rm p}(x)&=&
- {\sum_{i}}' s_i\xi_i 
\theta_i^\ih (x) 
\nonumber\\
&&\hspace{-1cm}
- p_\ih [u_N\delta(x-x_N^\ih)-u_1\delta (x-x_1^\ih)],
\ena 
where $\theta_i^\ih (x) 
=\theta (x-x_i^\ih) -
\theta (x,x_{i+1}^\ih)$. The 
last term in Eq.(A7)  vanishes  in the 
range $x_1^\ih<x<x_N^\ih$. 
We  then consider   the   correlation function 
for the thermal fluctuations of $\delta \Pi^{\rm p}$ defined by    
\be 
C(x,x')= \av{\delta \Pi^{\rm p}(x)\delta \Pi^{\rm p}(x')}_\ih/k_BT
\en  
where   $x$ and $x'$ are  in the range  $[x_1^\ih, x_N^\ih]$. 
Since $P_{\rm ih}$ is Gaussian, some calculations readily give  
\be 
{C (x,x')} = {\sum_{i}}'  s_i  
\theta_i^\ih (x) 
\theta_i^\ih (x') -\frac{K}{H}.
\en 
The  first term is   nonvanishing 
only when $x$ and $x'$ are in the same interval, 
so it is  short-ranged. The  second term 
is a constant ($\propto H^{-1})$, which arises from the global 
elastic coupling.  As will be shown in  Eqs.(A14) and (A16), 
$K$ has the meaning of the elastic constant  given  by 
\be
K = H/[{\sum_{i}}' s_i^{-1} + 2s_0^{-1}] = [H/(N+1)]/\av{s^{-1}}_\ih,
\en 
where $\av{s^{-1}}_\ih$ is 
the average of  $s_i^{-1}$ over all the bonds.

 To derive   Eq.(A9)  we can  use     the 
variance relations, 
\bea 
&&\hspace{-1cm}
\av{\xi_i\xi_j}_\ih= c_i\delta_{ij} -c_i e_j,~ 
\av{u_1^2}_\ih=\av{u_N^2}_\ih= c_0-c_0e_0,
 \nonumber\\  
&&\hspace{-1cm}
\av{\xi_i u_N}_\ih=-\av{\xi_i u_1}_\ih=  c_i e_{0} ,
\ena  
where   $c_i=k_BT/s_i$ and $e_i= K/Hs_i$.
We obtain the displacements  
from  $u_i=u_0+ \sum_{1\le k<i}\xi_k$. Thus\cite{Luben},   
\be 
\av{(u_i- u_j)^2 }_\ih/ k_BT= \ell_{ij}-K \ell_{ij}^2/H  , 
\en 
 where $\ell_{ij}=  \sum_{i\le k<j} 1/s_k$  for $i<j$. 
The two terms in Eq.(A12) grow with  increasing $j-i$  
 but  largely cancel for $\ell_{ij} >H/2K$. 
 For $i=1$ and $j=N$, they nearly cancel as   
$\av{(u_N- u_1)^2}_\ih/k_BT= 2(1-2e_0)/s_0$. 
We also find    the counterpart  of Eq.(50) in the form,  
\be 
s_0^2 \av{ u_1 u_N}_\ih/k_BT= K/H. 
\en


Next, we  shift the top  as   $R_N \to R_N+ \delta R_N$ 
keeping  the bottom at rest with mean strain 
 $\epsilon_{\rm ex}= \delta R_N/H $. 
The potential energy $U$  changes  by 
$ -p_\ih \delta R_N +\delta U'$ with 
\bea 
&&\hspace{-1.1cm} 
\delta U'=  \frac{1}{2} {\sum_{i}}'{s_i} { \xi_i^2 } 
+  \frac{1}{2} {s_0}  u_1^2+  
 \frac{1}{2}  {s_0} (u_N-\delta R_N)^2\nonumber\\
&&=K \epsilon_{\rm ex}^2H/2,  
\ena  
which  corresponds to Eq.(46). 
 Minimization of the first line  of Eq.(A14) yields 
 shifts of the particle   positions 
given by 
$
\xi_i=K {\epsilon_{\rm ex}}/{s_i}$, $
u_1= \delta R_N-u_N= K {\epsilon_{\rm ex}}/{s_0}$,  
for which   we obtain the second line.

We   also use  the linear response theory\cite{Kubo}  
 for  a small strain. 
The  perturbed  Hamiltonian is  
$
{\cal H}_\ih'= 
\sum_i |{\bi p}_i|^2/2m_i +\delta U -\epsilon_{\rm ex} \delta{\cal A}
$, where the first term is the kinetic 
energy, $\delta U$ is given by Eq.(A5), and 
\be
 \delta{\cal A}=
 \frac{s_0}{2}H  (u_N-u_1) 
= \int d{x}
\delta\Pi^{\rm p}-\sum_i x_i^\ih f_i, 
\en  
where  $f_i= - \p (\delta U)/\p u_i$.
As in Eq.(38) we obtain  
\be  
K=- \int dx \av{\delta \Pi^{\rm p}(x) \delta{\cal A}}_\ih/Hk_BT 
=K_{\rm a} - K_{\rm c}.
\en   
From Eq.(A15),     $K_{\rm a}$ is 
 the affine part and  $K_{\rm c}$ is 
the  nonaffine  part. From Eqs.(A7) and (A8) we find  
\bea 
&& 
\hspace{-1.2cm} 
 K_{\rm a}= - {\sum_i}' \frac{x_i^\ih}{H}
 \frac{\p}{\p u_i} \int dx 
\delta\Pi^{\rm p}= \frac{1}{H} {\sum_i}' s_i (d_i^\ih)^2, \\
&& \hspace{-1.2cm} 
K_{\rm c}= \frac{1}{H}
\int dx \int dx'  { C (x,x')} =K_{\rm a}-K . 
\ena 
In $K_{\rm c}$,  the  first  and  second  terms  in Eq.(A9)  
yield  $K_{\rm a}$ and $-K$, respectively, after  double integration.

 We notice that, if some bonds are very weak (with very small $s_i$), 
they can dominantly  contribute to  $\av{s^{-1}}_\ih$ 
giving rise to a large reduction in  $K$. In such cases, 
 $K$ can be much smaller than $ K_{\rm a}$.

\vspace{2mm}
\noindent{\bf Appendix B: Thermal fluctuations in solid films 
in continuum elasticity}\\
\setcounter{equation}{0}
\renewcommand{\theequation}{B\arabic{equation}}
  
In the  linear elasticity, 
 we consider   a solid film 
with  homogeneous elastic moduli, 
where the displacement field ${\bi u}({\bi r})$ 
is well-defined.  We assume that the  thermal fluctuations of 
$\bi u$  obey the distribution $\propto \exp(- F_{\rm el}/k_BT)$, where 
 $F_{\rm el}$ is  the elastic free energy\cite{Gusev,Binder,Rahman}. 
Here, the average over this distribution is written as $\av{\cdots}$. 
We impose  the rigid boundary condition 
at $z=\pm  H/2$ and the periodic boundary condition 
along the $x$ and $y$ axes with period $L$. 
The film volume is $V=HS$ with $S=L^{d-1}$. 
.

First,  we examine  the equal-time correlations. For simplicity, we  consider  
the   lateral average of $u_x$ (the zero wavenumber 
component in the $x$-$y$ plane) given by 
\be 
U_x(z) = S^{-1} \int d{\bi r}_\perp u_x({\bi r}), 
\en 
where $\int d{\bi r}_\perp$ is the integral  in the $x$-$y$ plane.
The normalized eigenfunctions 
  are $e_n(z)= \sqrt{2/V}\sin [k_n(z+H/2)] $ $(n\ge 1$)  
with $k_n= \pi n/H$ for  $U_x(\pm H/2)=0$.  As in Eq.(55), we introduce the 
 fluctuating variables $s_n$ by  
\be 
U_x(z) = \sum_{n\ge 1}   e_n(z)s_n.
\en  
As in Eq.(57), the   elastic free energy is expressed as    
 \be 
 F_{\rm el}=  \frac{1}{2}\mu S
 \int dz 
( U_x')^2
= \frac{ 1 }{2}\mu \sum_{n\ge 1} k_n^2  s_n^2 , 
\en 
where $\mu$ is the shear modulus 
and  $U_x'(z) =dU_x(z)/dz$. Then, we find 
$\av{s_n s_\ell}= k_BT \delta_{n\ell}/\mu k_n^2$   
 and     
\bea 
&&\hspace{-1.6cm}
\av{U_x(z) U_x(z')}/{k_BT }= 
 \sum_{n \ge 1} 
e_n(z) e_n(z')/{\mu k_n^2} \nonumber\\
&&\hspace{0.7cm}
= [{H^2}- 2{H}|z-z'|  -4zz']/{4\mu V},
\ena 
where   we  use  the formula  
 $\sum_{n\ge 1}\sin( n p)\sin( n q)/n^2=
 q(\pi-p)/2$ for $0<q<p<\pi$.  Differentiation of Eq.(B4) 
with respect to $z$ and $z'$ gives     the strain correlation, 
\be 
\av{U_x'(z) U_x'(z')}/ {k_BT}= [H \delta(z-z')  -1]/\mu V. 
\en  
Here, the $\delta$-function appears, but 
it should be regarded as a function with a microscopic 
width in particle  systems.    

In Eq.(28) we have introduced the forces  from the  walls. 
In the  continuum theory, we express them as  
\be 
{\delta F_{\rm top}^x}={S}\mu
U_x'({H'}/{2}),\quad   
{\delta F_{\rm bot}^x}=-{ S}\mu U_x'(-H'/2).
\en 
To account for the particle discreteness, 
we  assume  the stress  balance 
at $z=\pm H'/2$, 
where  $H'= H-\ell_{\rm m}$  with $\ell_{\rm m}$ being a microscopic length. 
For $|z|< H'/2$, 
we    find  $\av{U_x(z)\delta F_{\rm top}^x}/k_BT 
=-z/H-1/2$ and  $\av{U_x(z)\delta F_{\rm bot}^x}/k_BT
=z/H-1/2$ from Eq.(B4). These  lead to  the 
counterparts of Eqs.(35) and  (47). If we define  $\delta{\cal A}=   H
(\delta F_{\rm bot}^x- \delta F_{\rm top}^x)/2$ as in Eq.(27), we find 
\be
\av{U_x(z)\delta {\cal A}}/ k_BT =z  ,
\en   
which consists of     the affine displacement only.
To be precise,  $\av{U_x(z)\delta {\cal A}}$ nearly 
vanishes in the narrow layers  $H'/2<|z|<H/2$. If we 
assume the linear combination  
  $\delta {\cal A}=\sum_n Z_A^{(n)} s_n$ in terms of $s_n$ in Eq.(B2), 
 the coefficients $Z_A^{(n)}$ 
are nonvanishing only for even positive $n$ as 
 \be 
Z_A^{(n)} = -2\pi (\sqrt{2V}/H)\mu n   \quad ({n=2,4,\dots}). 
\en

Next, we examine  the time-correlations 
assuming the wave equation $\p^2 U_x/\p t^2= c_\perp^2 \p^2 U_x/\p z^2$ 
without dissipation, where $c_\perp$ 
is the transverse sound speed. Fixing $z'$ in the cell  $|z'|<H/2$, we have  
\be 
\av{U_x(z,t)U_x(z',0)}
=  g(z+c_\perp t, z')
+ g(z-c_\perp t,z'),
\en  
where $2g(z,z')$  is equal to  Eq.(B4) for $|z|<H/2$. 
We  extend it outside the cell setting   
 $g(z,z')= -g(H-z,z')= g(z+2H,z')$. We can obtain Eq.(B9) 
 if we replace  $e_n(z) e_n(z')$  by 
$e_n(z) e_n(z')\cos(c_\perp k_n t)$ in the first line of  Eq.(B4). 
 Then, Eq.(B9) gives  a periodic 
 function of $t$ and the period 
is twice larger than  the acoustic 
 traversal time $t_{\rm a}=H/c_\perp$. From   Eqs.(B6) and (B9), 
the function ${\cal F}_{\rm w}(t)$ defined in Eq.(82) is 
calculated as       
\be
{\cal F}_{\rm w}(t)= \mu-  \mu\sum_{\ell=0,\pm 1,\pm 2, \cdots }
 \delta (t/t_{\rm a} - 1-2\ell), 
\en
where the second term arises from  impulses due to repeated reflections  
of  transverse sounds without scattering. Thus, 
Eq.(B10)  is   consistent   with    Fig.8 for $t<t_{\rm a}$.

\end{document}